\documentclass[11pt,a4paper]{article} \usepackage{epsfig}
\usepackage{latexsym}
\usepackage{cite}
\usepackage{graphicx}
\newcommand{\lo}{\langle} \newcommand{\rc}{\rangle}
\newcommand{\ud}{{\mathrm d}}
\begin{document}
\title{Dynamical Rare event simulation techniques for equilibrium and non-equilibrium systems}
\author{Titus S. van Erp\\
Centrum voor Oppervlaktechemie
en Katalyse, K.U. Leuven,\\ Kasteelpark Arenberg 23, B-3001 Leuven, Belgium\\
E-Mail: {\tt Titus.VanErp@biw.kuleuven.be}}
\date{December 2010}
\maketitle

\tableofcontents

\section{Introduction}
Molecular dynamics (MD) is the ultimate method  to  
gain detailed atomistic information of dynamical
processes that are  difficult to access experimentally.
However, an important bottleneck of atomistic simulations 
is  the limited system- and timescales. 
Depending on the complexity of the forcefields (Ab Initio 
MD being extremely more expensive than classical MD) systems typically consist
of
100 to 100000 
molecules that can be simulated for a period of nanoseconds till microseconds.  
Therefore, many activated processes can not be studied using 
brute-force  MD 
because the probability to observe a reactive event within reasonable CPU time is basically zero.
Typical examples are
protein folding, conformational changes of molecules, cluster isomerizations, chemical reactions, diffusion in solids, ion permeation through membranes, enzymatic reactions, docking, nucleation, DNA denaturation, and other types of phase transitions.
If these processes are treated with straightforward MD, the simulations 
will endlessly remain in the reactant states.
Still, if  an event would happen, it can go very fast.  
The time it actually takes to cross the 
barrier
is usually much shorter than this computational accessible 
timescale.  
Therefore, rare event  algorithms aim to avoid the superfluous exploration of the reactant
state and to enhance the occurrence of reactive events. 
The methods that I will discuss are the 
reactive flux (RF) method~\cite{Chandlerbook,FrenkelSmit} and the more 
recent algorithms that originate from the 
transition path sampling (TPS)~\cite{TPS98,TPS98_2,TPS99,Bolhuis02,Dellago02} 
methodology. These comprise the transition interface sampling 
(TIS)~\cite{ErpMoBol2003} and the replica 
exchange TIS (RETIS)~\cite{vanErp07PRL,RETIS2}, which are successive improvements 
on the way reaction rates were determined in the original TPS algorithm.
Partial path TIS (PPTIS)~\cite{MoBolErp2004} is an approximative approach 
in order to reduce the simulated path length  for the case of diffusive barrier crossings. 
PPTIS is similar to Milestoning~\cite{FaEl2004}, that was developed simultaneously and independently from PPTIS.  For non-equilibrium systems, the Forward Flux Sampling (FFS) was designed~\cite{FFS}. This method  
is based on the TIS formalism, but does not require prior knowledge on the phasepoint density.
All these methods have in common that they aim to simulate true
molecular dynamics trajectories at a
much faster rate than naive brute force molecular dynamics.
I will discuss the advantages and disadvantages of the different methodologies
and introduce a few new relations and derive some known relations using a 
nonstandard approach. The descriptions of these methods given here are 
far from complete and,
therefore,  to obtain a more complete picture of the path sampling techniques 
I would like to recommend some very recent complementary reviews
on these
methodologies,\cite{revTPS2009,FFSrev2,FFSrev,revTPS2010}.
In the end, I compare all the methods by applying them on a simple, though tricky, test system. The outcome 
illustrates some important pitfalls for the non-equilibrium methods 
that have no easy solution and show that
caution is necessary when interpreting  their results.

\section{Reactive Flux Method}
Low dimensional systems, such as chemical reactions in the gas phase, 
are usually well described by 
Transition State Theory (TST). TST assumes that
the transitions from reactant to product state always follow a path on the potential energy surface
such that it passes the barrier 
nearby the 
transition state (TS). The TS  refers to the point on the  
energy barrier
having to the lowest possible potential energy difference, with respect to the 
reactant state, that any trajectory must overcome in order to reach  the 
product state. In this description, 
TS corresponds to a unstable stationary point 
on the potential energy
surface having one imaginary frequency (saddle point). In condensed  systems,
the saddle-point of the full potential energy surface is usually less  meaningful.
For instance, if we would consider the dissociation of NaCl in water, the TS would correspond
to a state where the inter-ion distance is fixed to a critical value while all surrounding water molecules are full frozen into an icy state. It is needless to say that this does  not 
correspond to our daily experience
when dissolving some pinch of salt in a glass of water.
 TST can be generalized for  higher dimensional systems using the 
free energy instead of the potential energy. The TS
is then no longer a single point, but a multidimensional surface.
In this case, the TST equation is determined by the free energy difference 
between the TS dividing surface and the reactant state. An important 
limitation of TST is that the free energy barrier depends on the selected
degrees of freedom that are used to describe the free energy surface.
In addition, TST implicitly assumes that any trajectory will cross 
the free energy barrier only once when going from reactant to product state.
Kramers' theory\cite{50Kramer} provides an elegant and insightful 
approach to correct for correlated recrossings if these originate
from the diffusive character of the dynamics. 
However, there are several other sources 
for recrossings.  For instance, if the 
selected degrees of freedom are not well chosen, the barriers in the free energy landscape
do not always correspond the barriers in the  underlying potential energy surface which ultimately determine the 
dynamics.
The reactive flux (RF) method is able to correct for recrossings
regardless their origin and
is very powerful when 
when TST is 
close but not sufficiently accurate.

The theory of the method 
originated from the
early 1930s,  far before the first applications of computers for molecular dynamics 
simulations~\cite{Alder}.  
After Wigner and Eyring 
introduce the concept of the TS
and the TST approximation~\cite{E35,W38},  
Keck \cite{Keck62} demonstrated how to calculate the dynamical
correction, the transmission coefficient. 
This work has later been extended by
Bennett~\cite{Bennet77}, Chandler~\cite{DC78} and
others~\cite{Yamamoto60,H38}, resulting in a two-step approach. First
the free energy as function of a single reaction coordinate (RC) is
determined. This can be done by e.g. umbrella sampling (US)
\cite{TV74} or thermodynamic integration (TI)~\cite{CCH89}.  Then, the
maximum of this free energy profile defines the approximate TS
dividing surface and the transmission coefficient can be calculated by
releasing dynamical trajectories from the top.  

Traditionally, the equation for the dynamically corrected rate constant 
is derived by
applying a small perturbation to the equilibrium state and invoking
the fluctuation-dissipation theorem and Onsager's 
relation~\cite{Yamamoto60,Chandlerbook,FrenkelSmit}.
However, as I will show here, there is an alternative derivation 
that naturally evolves to a formula 
for transmission coefficient
that is probably more efficient that the standard one \cite{Bennet77,DC78}. 

There are several definitions for the rate constant $k_{AB}$ between two 
states $A$ and $B$, such as the 
transition probability per unit time, the inverse mean residence time 
in state $A$, or the inverse mean
first passage time towards state $B$~\cite{Moronithesis}.  
However, all these different definitions
become equivalent for truly exponential relaxation, which is the case whenever 
the stable states $A$ and $B$ are separated by large free energy barriers. If this is not the case,
the rate constant becomes ill-defined.
To start the derivation I will use the first definition, which can be expressed as follows:
\begin{eqnarray}
k_{AB}=\lim_{\ud t \rightarrow 0}
\frac{1}{\ud t}
 \frac{\textrm{number of states $A$ that transform into state $B$ within $\ud t$}}
{\textrm{number of states $A$}}
\label{kphenom}
\end{eqnarray}
Let us denote $x=(r,v)$ the phasepoint which includes the positions $r$ and velocities $v$ of
all particles in the system.
We define the reaction coordinate $\lambda(x)$ which can be any function of $x$, though in practice it will generally only depend on $r$.  The RC function should describe the 
progress of the reaction, but there is a lot of flexibility 
in designing this RC function.

We will assume that
the collection of phasepoints $\{ x|\lambda(x)=0\}$ defines the transition state dividing surface that 
separates region $A$ and $B$. For convenience, we will also
assume that the RC will increase when going from $A$ to $B$.
Considering the phenomenological equation~\ref{kphenom}, we can directly
write down the reaction rate as
\begin{eqnarray}
k_{AB}= \lim_{\ud t \rightarrow 0}\frac{1}{\ud t} \frac{
\int \ud x_0  \, \theta\left(-\lambda\left(x_0 \right)\right) \theta\left( \lambda\left(x_{\ud t}\right) \right) \rho(x_0) 
}{
\int \ud x_0  \theta\left(-\lambda\left(x_0 \right)\right) \rho(x_0) 
}=
\lim_{\ud t \rightarrow 0}
\frac{1}{\ud t}
\frac{\lo   \theta\left(-\lambda\left(x_0 \right)\right) \theta\left( \lambda\left(x_{\ud t}\right) \right)  \rc}{\lo
 \theta\left(-\lambda\left(x_0 \right)\right) 
 \rc}
 \label{TST0}
\end{eqnarray}
where $x_0$ and $x_{\ud t}$ are phasepoints at times $t=0$ and $t=\ud t$.
$\rho(x)$ denotes the phasepoint density. For equilibrium statistics this is simply given by Boltzmann $\rho(x)=\exp(-\beta E(x))$ where $E$ the energy and $\beta=1/k_B T$, $T$ the temperature, and $k_B$ the Boltzmann constant. $\theta$ is the Heaviside-step function
with $\theta(y)=0$ if $y<0$ and  $\theta(y)=1$ otherwise. The brackets
$\lo \ldots \rc \equiv \int \ud x \, \dots \rho(x)/ \int \ud x \,  \rho(x)$ 
denote the ensemble average over the
initial condition $x_0$. Eq.~\ref{TST0} is basically the TST expression of the rate, but written
in a somewhat unusual form. 

To transform this equation into the standard form, we can use
$\lambda\left(x_{\ud t}\right)= \lambda\left(x_{0}\right)+\ud t \dot{\lambda}(x_0)+
{\mathcal O}(\ud t^2)$, where the dot denotes the time derivative.
If we neglect the second order terms, we can write for an arbitrary function   $a(x)$: 
\begin{eqnarray}
\int \ud x_0 \,  \theta\left(-\lambda\left(x_0 \right)\right) \theta\left( \lambda\left(x_{\ud t}\right) \right) a(x_0)= 
\int \ud x_0 \,  \theta\left(-\lambda\left(x_0 \right)\right) \theta\left( \lambda\left(x_{0}\right)+
\ud t \dot{\lambda}\left(x_0\right) \right)  a(x_0) 
\label{p1sub1}
\end{eqnarray}
Clearly, $\theta(-\lambda ) \theta( \lambda+
\ud t \dot{\lambda} )$ is only nonzero if $\dot{\lambda}>0$ and
$\ud t \dot{\lambda} <\lambda<0$. Instead of integrating over $x_0$, we will apply a coordinate transform such that we can integrate Eq.~\ref{p1sub1} over $\lambda, \dot{\lambda}$ and a
remaining set coordinates $x_0'$.  Assume that  $J(x_0',\lambda,\dot{\lambda})$ is the corresponding Jacobian of this transformation.  We can then integrate out the $( \lambda, \dot{\lambda})$ coordinates
\begin{eqnarray}
&&
\int \ud x_0 \,  \theta\left(-\lambda\left(x_0 \right)\right) \theta\left( \lambda\left(x_{0}\right)+
\ud t \dot{\lambda}\left(x_0\right) \right)  a(x_0) =
\int \ud x_0' \int_0^\infty \ud \dot{\lambda} \int_{-\ud t \dot{\lambda}}^0 \ud \lambda \,
    a(x_0',\lambda,\dot{\lambda}) J(x_0',\lambda,\dot{\lambda})  \nonumber \\
&=& \int \ud x_0' \int_0^\infty \ud \dot{\lambda} \int_{-\ud t \dot{\lambda}}^0 \ud \lambda \,
     \Big\{ a(x_0',0,\dot{\lambda}) J(x_0',0,\dot{\lambda})  
     +\lambda \frac{\partial (a(x_0',\lambda,\dot{\lambda}) J(x_0',\lambda,\dot{\lambda}) )}{\partial        
      \lambda}\Big|_{\lambda=0} + \ldots\Big\} \nonumber \\ 
&=&\int \ud x_0' \int_0^\infty \ud \dot{\lambda} \,
      \Big\{  (\ud t \dot{\lambda})
      a(x_0',0,\dot{\lambda}) J(x_0',0,\dot{\lambda})   -
        \frac{1}{2} (\ud t \dot{\lambda})^2
        \frac{\partial (a(x_0',\lambda,\dot{\lambda}) J(x_0',\lambda,\dot{\lambda}) )}{\partial \lambda}
        \Big|_  {\lambda=0}   + \ldots\Big\}   \nonumber \\
  &=& \ud t  \times \int \ud x_0' \int_0^\infty \ud \dot{\lambda} \,
     \dot{\lambda}
     a(x_0',0,\dot{\lambda}) J(x_0',0,\dot{\lambda})   
      + \mathcal{O}(\ud t^2)
 \end{eqnarray}
 where we applied a Taylor expansion in terms of $\lambda$ in the second line.  
 Clearly, as 
 \begin{eqnarray}
  \int \ud x_0 \, 
  \dot{\lambda}\left(x_0\right) 
  \delta\left(\lambda\left(x_0\right)  \right) 
  \theta\left(\dot{\lambda}\left(x_0\right)  \right) a(x_0)  &=&
  \int \ud x_0'   \ud \dot{\lambda} \ud \lambda \,   \dot{\lambda}
  \delta(\lambda)  
  \theta(\dot{\lambda}) 
    a(x_0',\lambda,\dot{\lambda}) J(x_0',\lambda,\dot{\lambda})  \nonumber \\
  &=& \int \ud x_0' \int_0^\infty \ud \dot{\lambda} \,
   \dot{\lambda}
  a(x_0',0,\dot{\lambda}) J(x_0',0,\dot{\lambda}) 
\end{eqnarray} 
we have proven that 
\begin{eqnarray}
\lim_{\ud t \rightarrow 0} \frac{1}{\ud t } \theta\left(-\lambda\left(x_0 \right)\right) \theta\left( \lambda\left(x_{\ud t}\right) \right)  =   \dot{\lambda}\left(x_0\right) 
  \delta\left(\lambda\left(x_0\right)  \right) 
  \theta\left(\dot{\lambda}\left(x_0\right)  \right)
  \label{ttd}
\end{eqnarray}
Using this expressing into Eq.~\ref{TST0}, we obtain the standard form of the {\bf TST formula}:
\begin{eqnarray}
k_{AB}=
\frac{\lo  
  \dot{\lambda}\left(x_0\right) 
  \delta\left(\lambda\left(x_0\right)  \right) 
  \theta\left(\dot{\lambda}\left(x_0\right)  \right)
  \rc}{\lo
 \theta\left(-\lambda\left(x_0 \right)\right) 
 \rc}
 \label{TST}
\end{eqnarray}
It is often very convenient to switch back to the other formalism, Eq.~\ref{TST0}, 
as some relations follow
 more naturally from this expression, especially in path sampling 
simulations where $\ud t$ can  simply be taken as the MD timestep.  
The TST approach 
rewrites Eq.~\ref{TST} into two factors
\begin{eqnarray}
k_{AB}=
\frac{\lo  
  \dot{\lambda}\left(x_0\right) 
  \delta\left(\lambda\left(x_0\right)  \right) 
  \theta\left(\dot{\lambda}\left(x_0\right)  \right)
  \rc}{\lo  \delta\left(\lambda\left(x_0\right)  \right)   \rc} \times
  \frac{\lo   \delta\left(\lambda\left(x_0\right)  \right)  \rc}{\lo
 \theta\left(-\lambda\left(x_0 \right)\right) \rc} \equiv R^{\rm TST}   \times
 \frac{e^{-\beta F(0)}}{\int_{-\infty}^0  \ud \lambda \, e^{-\beta F(\lambda)}}
\end{eqnarray}
where the free energy $F$ is defined as $F(\lambda)\equiv-\ln  \lo 
\delta(\lambda-\lambda(x)) \rc /\beta$.
Numerous techniques exist to calculate the free energy profile along 
the barrier region~\cite{TV74,CCH89,WHAM,Briels,flood,Darve,WL,LP02}. 
The kinetic term
$R^{\rm TST}$ usually follows from a simple numerical or analytical integration.
For instance, if the RC is  a simple Cartesian coordinate of a target particle then
$R^{\rm TST}=1/\sqrt{2 \pi \beta m}$ where $m$ is the mass of the particle.

The TST expression neglects correlated fast recrossings and, therefore, overestimates
the reaction rate. Recrossings  can occur due to 
a diffusive motion on top of the barrier or by kinetic correlations when the kinetic energy of the 
RC is not dissipated. Another important source of recrossings is when the  one-dimensional RC
gives an incomplete description of the reaction kinetics~\cite{FrenkelSmit}. 
To correct for recrossing  we can apply  the effective positive flux formalism
which neglects the crossings that are not "effective". At each side 
of the barrier we define
regions that are the stable regions $A$ and $B$. 
These might be smaller than the regions that we associate
to the product and reactant state.
Entering $A$ or $B$ implies that the system
is committed to that side, i.e. it might leave region $A$ or $B$ shortly 
thereafter, but the chance to rapidly recross the barrier is of the same order 
as an independent new event. An effective positive crossing is then defined as the first crossing on the trajectory that makes the transition from $A$
to $B$ (See fig.~\ref{figepf}).  This leads to the {\bf effective positive flux}
expression for the reaction rate:
\begin{figure}[ht!]
  \begin{center}
  \includegraphics[width=11.1cm]{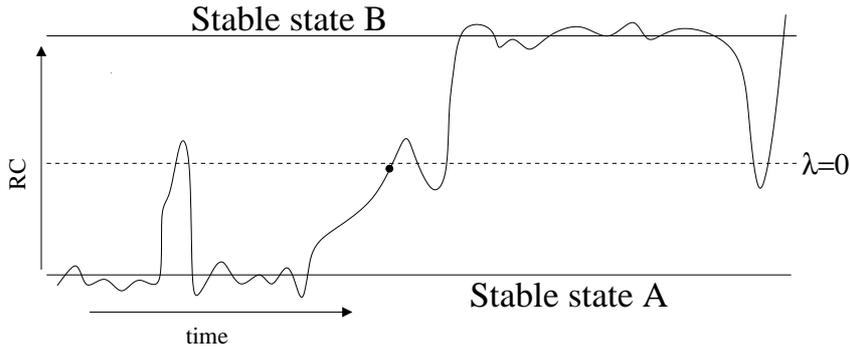}
   \caption{Definition  of an effective positive crossing
on a very long MD trajectory.
The EPF algorithm will ignore all crossing with the TS dividing surface except one (black dot).
These are the first crossing points with the TS dividing surface 
for the parts 
of the MD trajectory that start at $A$ and end at $B$ 
(without revisiting $A$ again).
\label{figepf}}
  \end{center}
\end{figure}
\begin{eqnarray}
k_{AB}&=&
\lim_{\ud t \rightarrow 0}
\frac{1}{\ud t}
\frac{\lo   \theta\left(-\lambda\left(x_0 \right)\right) \theta\left( \lambda\left(x_{\ud t}\right) \right)
h^b_{A0}(x_0) h^f_{BA} \left(x_0 \right)
  \rc}{\lo
 \theta\left(-\lambda\left(x_0 \right)\right) 
 \rc} \nonumber \\
 &=&
 \frac{\lo 
  \dot{\lambda}\left(x_0\right) 
  \delta\left(\lambda\left(x_0\right)  \right) 
  \theta\left(\dot{\lambda}\left(x_0\right)  \right)
h^b_{A0}(x_0) h^f_{BA} \left(x_0 \right)
  \rc}{\lo
 \theta\left(-\lambda\left(x_0 \right)\right) 
 \rc}
 \label{kepf}
\end{eqnarray}
where $h^{b/f}_{uv}(x_0)$ detects whether a backward/forward time trajectory  crosses  or enters 
a certain interface/region $u$ before interface/region $v$. 
If this is true, the function is one. It is zero otherwise.  
In the second line we applied again equality~\ref{ttd}. Naturally,
the ensemble average $\lo \ldots \rc$ should now not only integrate over the phasepoint $x_0$,
but also sum over all possible trajectories backward and forward in time starting from $x_0$.
The ratio between the exact expression (Eq.~\ref{kepf}) and the transition state expression (Eq.~\ref{TST}) is the transmission coefficient: $k^{\rm exact}=\kappa k^{\rm TST}$ so that
\begin{eqnarray}
\kappa&=&
 \frac{\lo 
  \dot{\lambda}\left(x_0\right) 
  \delta\left(\lambda\left(x_0\right)  \right) 
  \theta\left(\dot{\lambda}\left(x_0\right)  \right)
h^b_{A0}(x_0) h^f_{BA} \left(x_0 \right)
  \rc}{\lo
\dot{\lambda}\left(x_0\right) 
  \delta\left(\lambda\left(x_0\right)  \right) 
  \theta\left(\dot{\lambda}\left(x_0\right)  \right)
 \rc}\nonumber \\
 &=&
  \frac{\lo 
  \dot{\lambda}\left(x_0\right) 
  \theta\left(\dot{\lambda}\left(x_0\right)  \right)
h^b_{A0}(x_0) h^f_{BA} \left(x_0 \right)
  \rc_{\lambda=0}}{\lo
\dot{\lambda}\left(x_0\right) 
  \theta\left(\dot{\lambda}\left(x_0\right)  \right)
 \rc_{\lambda=0}}
\label{kappaepf}
\end{eqnarray} 
Here the subscript $\lambda=0$ denotes an ensemble average on the TST dividing surface. Strictly speaking, the above expression is correct for any surface that separates the two stable states. However, the efficiency to calculate the 
above expression is significant better if $\kappa$ is maximized. 
Therefore, $\lambda=0$ should be defined on the top of the free energy barrier.
If we assume that $\lambda(x)=\lambda(r)$ depends on configuration space only,
the calculation requires to generate a representative set 
of configuration points
on the TST surface.
Then, we attribute to these points $r$ a randomized set of velocities 
taken from a  Maxwellian distribution and integrate the equations of motion backward and forward in time. However, as 
$ \theta\left(\dot{\lambda}  \right)
 h^b_{A0} =0$ if $\dot{\lambda} <0$  or when the backward trajectory recrosses 
the TST dividing surface before entering $A$, only a very few trajectories 
need to be fully integrated  in both time-directions until reaching stable states.
It is surprising that the effective
positive flux  counting strategy is not so common.  To our knowledge only two slightly different
expressions of a transmission coefficient based on the effective
positive flux have been proposed in
Refs.~\cite{Anderson95,titusthesis}.
All other expressions in the literature do not avoid 
the counting of recrossings. In these algorithms, the final rate
constant follows through cancellation of many negative and positive
terms.  
For instance, the most popular formulation of the rate constant and transmission coefficient is the {\bf Bennett-Chandler (BC)} expression that
appears in many textbooks on molecular 
simulation~\cite{Chandlerbook,FrenkelSmit}. 
\begin{eqnarray}
\tilde{k}_{AB}(t)&=&
 \frac{\lo 
  \dot{\lambda}\left(x_0\right) 
  \delta\left(\lambda\left(x_0\right)  \right) 
 \theta\left(\lambda\left(x_t \right)\right)
  \rc}{\lo
 \theta\left(-\lambda\left(x_0 \right)\right) 
 \rc} \Rightarrow \nonumber \\
\tilde{\kappa}(t)&=&
  \frac{\lo 
  \dot{\lambda}\left(x_0\right) 
\theta\left(\lambda\left(x_t \right)\right)
  \rc_{\lambda=0}}{\lo
\dot{\lambda}\left(x_0\right) 
  \theta\left(\dot{\lambda}\left(x_0\right)  \right)
 \rc_{\lambda=0}}
 \label{BC}
\end{eqnarray}
Here, the reaction rate and transmission coefficient are 
expressed as time-dependent functions. However, the actual rate 
constant and transmission coefficient, which should not depend on time, 
follow from a plateau value of these time-dependent functions: $k_{AB}=\tilde{k}_{AB}(t'), \kappa=\tilde{\kappa}(t')$ with $\tau_{\rm mol} < t' <\tau_{\rm rxn}$. In other words, $\tilde{k}(t)$ and 
$\kappa(t)$ will generally show oscillatory behavior at small $t$. However,
after some molecular timescale $\tau_{\rm mol}$, the system will basically enter either region $A$ or $B$ (See fig.~\ref{figepf}) after which we won't expect 
any recrossing until reaching the actual relaxation time $\tau_{\rm rxn} >> 
\tau_{\rm mol}$. The equivalence between Eq.~\ref{BC} and 
Eqs.~\ref{kepf}, \ref{kappaepf}, can be shown by invoking 
$\theta\left(\lambda\left(x_{t'} \right)\right)=h^f_{BA}(x_0)$,
Eq.~\ref{ttd}, and its mirror equivalent 
$ \lim_{dt\rightarrow 0}
\theta(\lambda(x_0)) \theta( -\lambda(x_{\ud t}) )/dt  = -\dot{\lambda}(x_0) 
  \delta(\lambda(x_0)) 
  \theta(-\dot{\lambda}(x_0))$:
\begin{eqnarray}
k&=& \tilde{k}(t')=
 \frac{\lo 
  \dot{\lambda}\left(x_0\right) 
  \delta\left(\lambda\left(x_0\right)  \right) 
 \theta\left(\lambda\left(x_{t'} \right)\right)
  \rc}{\lo
 \theta\left(-\lambda\left(x_0 \right)\right) 
 \rc}
=
\frac{\lo 
  \dot{\lambda}\left(x_0\right) 
  \delta\left(\lambda\left(x_0\right)  \right) 
h^f_{BA} \left(x_0 \right) 
  \rc}{\lo
 \theta\left(-\lambda\left(x_0 \right)\right) 
 \rc} 
\nonumber \\
&=&
\frac{\lo 
  \dot{\lambda}\left(x_0\right) 
  \delta\left(\lambda\left(x_0\right)  \right) 
 \theta\left(\dot{\lambda}\left(x_0\right)  \right)
h^f_{BA} \left(x_0 \right) 
+
  \dot{\lambda}\left(x_0\right) 
  \delta\left(\lambda\left(x_0\right)  \right) 
 \theta\left(-\dot{\lambda}\left(x_0\right)  \right)
h^f_{BA} \left(x_0 \right) 
  \rc}{\lo
 \theta\left(-\lambda\left(x_0 \right)\right) 
 \rc} 
\nonumber \\
&=&
\lim_{\ud t \rightarrow 0}
\frac{1}{\ud t} \left(
\frac{\lo   \theta\left(-\lambda\left(x_0 \right)\right) 
\theta\left( \lambda\left(x_{\ud t}\right) \right)
h^f_{BA} \left(x_0 \right)
-
\theta\left(\lambda\left(x_0 \right)\right) 
\theta\left(- \lambda\left(x_{\ud t}\right) \right)
h^f_{BA} \left(x_0 \right)
  \rc}{\lo
 \theta\left(-\lambda\left(x_0 \right)\right) 
 \rc} \right)
 \label{ttdBC}
\end{eqnarray}
We have now transferred the BC expression in an unitary 
ensemble average;
each phasepoint $x_0$ either returns 1, 0, or -1. Consider
a very long MD trajectory with a timestep of $\ud t$ (like the one in fig.~\ref{figepf}). It is clear
that any detailed-balance simulation method should sample each phasepoint
$x_0$ on this trajectory equally often. As such, 
an unreactive $B\rightarrow B$
trajectory will always have an equal number of phasepoints returning $+1$ as $-1$. The $B\rightarrow B$ trajectories are therefore effectively not counted due to this cancellation. The phasepoints on the $A\rightarrow A$ trajectory are always zero due to the $h^f_{BA}$ characteristic function. Finally,
any trajectory $A\rightarrow B$ always has one $x_0$ more that is +1 than 
-1.  A more formal mathematical proof of the equivalence between Eq.~\ref{BC} and Eq.~\ref{kepf} can be found in Ref.~\cite{EricvE}.

Whenever, there are a significant number of recrosssings, 
the BC formalism has obvious disadvantages. 
In general, we note that
any averaging
method  counting only zero and positive values will show a faster
convergence than  one that is based on  cancellation of positive
en negative terms. Moreover, in the effective flux formalism many
trajectories will be assigned as unreactive after just a few MD steps,
thus reducing the number of required force
evaluations.
Another important advantage of the EPF formalism is that is generates a set of 
trajectories that are unambiguous interpretable as reactive or unreactive,
while the BC scheme generates only forward trajectories of which some
actually belong to unreactive $B \rightarrow B$ trajectories.
Instead of integrating the equations of motion until reaching stable states,
one can also use a time-dependent expression for the EPF~\cite{TISeff}
similar to Eq.~\ref{BC}.

There are several other formulations of the transmission coefficient
(see~\cite{ErpBol2005}), but most of them rely on a cancellation between
positive and negative flux terms.
A comparative
study of ion channel diffusion \cite{White2000} showed that the
algorithm based on effective positive flux expression 
was superior to the other transmission rate
expressions. Moreover, it was as efficient as an optimized
version of the more complicated method of 
Ruiz-Montero {\emph et al.}\cite{Ruiz97}.
The implementation of the EPF scheme is as 
simple as algorithms that are based on
the BC transmission coefficient. Therefore, the EPF implementation of the RF
method should in principal be preferred above the standard implementations that
require cancellation.  

\section{Transition Path sampling}
In the previous section, I showed how the standard transmission coefficient 
calculations can be improved using the effective positive flux expression.
However, this approach can not fully eliminate the main bottleneck of the RF 
methods. If $\kappa << 1$ the number of trajectories that are required 
for sufficient statistics can be tremendous. In specific, if one is unable 
to find a proper RC, the overwhelming majority of trajectories that are 
released from the top of the barrier will be either 
$A \rightarrow A$ or  $B \rightarrow B$ trajectories~\cite{FrenkelSmit}.
In practice, it has been discovered that finding a good RC can be extremely difficult in high dimensional complex systems. Notable examples are chemical reactions in solution, where the reaction mechanism often depends on highly non- trivial solvent 
rearrangements~\cite{ErpMeij04}.  Also, computer simulations of nucleation processes use very complicated order parameters to distinguish between particles belonging to the liquid and solid phase. This makes it unfeasible to construct a single RC that accurately describes the exact place of cross-over transitions. As a result, hysteresis effects and low transmission coefficients are almost unavoidable.

This has been the main motivation of 
Chandler and collaborators~\cite{TPS98,TPS98_2,TPS99,Bolhuis02,Dellago02}
to devised a method that generates reactive trajectories without the need
of a RC. This  method, called transition path
sampling (TPS), gathers a collection of trajectories connecting the
reactant to the product stable region by employing a Monte Carlo (MC)
procedure called {\em shooting}. 

Suppose ${\bf x}$ is a path $\{x_0, x_{\ud t}, x_{2 \ud t}, \ldots,  
x_{n \ud t}\}$ of $n$ timeslices. The statistical weight given to this path 
equals
\begin{eqnarray}
P[{\bf x}]=\rho(x_0) p(x_0 \rightarrow x_{\ud t}) p(x_{\ud t}  \rightarrow x_{2 \ud t}) 
\ldots p(x_{(n-1) \ud t}  \rightarrow x_{n \ud t}) \hat{h}({\bf x})
\label{Pw}
\end{eqnarray}
where $\rho(x_0)$ is the usual phasepoint density and  
$p(x_{j\ud t}  \rightarrow x_{(j+1) \ud t})$ is probability 
density that the MD integrator generates  $x_{(j+1) \ud t}$ 
starting from $x_{j \ud t}$. The characteristic function $ \hat{h}({\bf x})$
equals 1 (otherwise 0) if a specific condition is fulfilled. For instance,
one could imply that the trajectory ${\bf x}$ needs to start in state $A$ and 
end in state $B$.

\begin{figure}[ht!]
  \begin{center}
  \includegraphics[width=11.1cm]{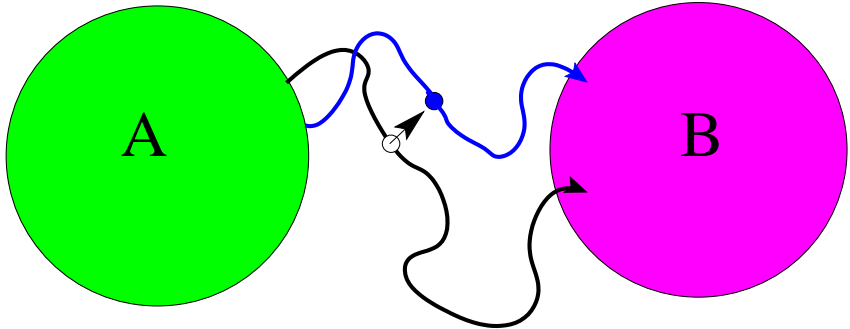}
   \caption{Illustration of the TPS shooting move using flexible path length.
From an existing path (black line) a random timeslice is selected. 
Positions and/or velocities of this point are slightly modified giving 
a new phasepoint (blue dot). From this point, the equations of motion are integrated forward and backward in time until the trajectory hits $A$ or $B$.
\label{figTPS}}
  \end{center}
\end{figure}

By means of the shooting algorithm, TPS performs a random walk in path space 
to generate one trajectory after the other
(See fig.~\ref{figTPS}).
The first step of this
approach consists of 
a random selection of one of the timeslices of the old path, called the shooting point. This timeslice
is modified by making random modifications in the velocities and/or positions.
Then, there is usually an acceptance or rejection 
step based on the energy difference between modified and unmodified shooting point. If accepted, the equations of motion are integrated forward and backward in 
time until a certain path length is obtained or until the condition function
$\hat{h}({\bf x})$ can be assigned 0 or 1. In the last case the trial move will be accepted. Any rejection along this scheme implies that the whole trial path
will be rejected and the old path is counted again just like in standard
Metropolis MC.
Naturally, the random walk in path space should obey  
detailed balance 
\begin{eqnarray}
\frac{P_\textrm{gen}[{\bf x}^{\rm (o)} \rightarrow {\bf x}^{\rm (n)}]}{ P_\textrm{gen}[{\bf x}^{\rm (n)} \rightarrow {\bf x}^{\rm (o)}]}
\frac{P_\textrm{acc}[{\bf x}^{\rm (o)} \rightarrow {\bf x}^{\rm (n)} ]}{ P_\textrm{acc}[{\bf x}^{\rm (n)} \rightarrow {\bf x}^{\rm (o)}]}=
\frac{P[{\bf x}^{\rm (n)}]}{P[{\bf x}^{\rm (o)}] },
\label{detbalpath}
\end{eqnarray}
where  the superscripts (o) and (n) denote the  old and new path respectively,
and
$P_{\rm gen}[{\bf x} \rightarrow {\bf x'}]$ is the probability to generate
path ${\bf x'}$ staring from ${\bf x}$.
Following the Metropolis-Hastings scheme, the acceptance rule
of the whole shooting move can  be written as
\begin{eqnarray}
P_{\rm acc}[{\bf x}^{\rm (o)} \rightarrow {\bf x}^{\rm (n)}]  = 
\hat{h}({\bf x}^{\rm (n)}) 
\min\left[ 1, \frac{P[{\bf x}^{\rm (n)}]}{P
[{\bf x}^{\rm (o)}]} \frac{P_{\rm gen}[{\bf x}^{\rm (n)} 
\rightarrow {\bf x}^{\rm (o)}]} 
{P_{\rm gen}[{\bf x}^{\rm (o)} \rightarrow {\bf x}^{\rm (n)}]}  \right].
\label{MH}
\end{eqnarray}

The generation probability is a product of different sub-probabilities.
These are $P_{\rm sel}$, to select the shooting point, $P_{\rm ran}$,
for the random modification of this shooting point, and $P_{\rm traj}$, which
is the probability to obtain ${\bf x}^{(n)}$ by integrating the equations
of motion backward and forward in time starting from the 
modified shooting point. 
\begin{eqnarray}
P_{\rm gen} [{\bf x} \rightarrow {\bf x'}]=P_{\rm sel} (x_{\rm shoot}|{\bf x})
\,
P_{\rm ran} (x_{\rm shoot} \rightarrow x'_{\rm shoot}) \,
P_{\rm traj}({\bf x'}| x'_{\rm shoot})
\end{eqnarray}
If we generate
paths of a fixed length $n$ and if each timeslice has an equal 
probability to be selected then
 $P_{\rm sel}=1/n$.  We come back to this point later on. 
In addition,  TPS algorithms generally utilize a symmetric random modification
of the shooting point: $P_{\rm ran} (x_{\rm shoot} \rightarrow x'_{\rm shoot})=
P_{\rm ran} (x'_{\rm shoot} \rightarrow x_{\rm shoot})$. 
Therefore, both $P_{\rm sel}$ and $P_{\rm ran}$ cancel out
in Eq.~\ref{MH}.
The acceptance rule simplifies even further
if we also  assume  
that the
dynamics obey
the microscopic reversibility condition
\begin{equation}
\label{equ:fw_bw}
\rho(x) p(x\rightarrow y)=\rho(y) p(\bar y \rightarrow
\bar x)
\label{micrev}
\end{equation}
where $\bar x$ is the phasepoint $x$ with reversed velocities: $\bar x=(r,-v)$.
This relation is very general and valid for a broad class of
dynamics applying to both equilibrium and non-equilibrium
systems~\cite{Dellago02}. 
By applying Eq.~\ref{micrev} several times on Eq.~\ref{Pw} we can show that
\begin{eqnarray}
P[{\bf x}]&=&\rho(x_{j \ud t}) 
p(\bar{x}_{j \ud t} \rightarrow \bar{x}_{(j-1) \ud t} ) 
p(\bar{x}_{(j-1) \ud t} \rightarrow \bar{x}_{(j-2) \ud t} ) \ldots
p(\bar{x}_{\ud t} \rightarrow \bar{x}_{0} ) \nonumber \\ 
&\times& p(x_{j \ud t} \rightarrow x_{(j+1) \ud t} ) 
p(x_{(j+1) \ud t} \rightarrow x_{(j+2) \ud t} ) \ldots
p(x_{(n-1) \ud t} \rightarrow x_{n \ud t} )
\label{Pw2}
\end{eqnarray}
is true for any timeslice $j$.
For time-reversible dynamics the
backward integration is simply obtained by reversing the velocities and integration forward in time.
Hence, 
the generation probability
$P_{\rm traj}({\bf x}| x_{\rm shoot})$ depends on exactly the same 
transition probabilities
$p(x\rightarrow x')$ and $p(\bar x \rightarrow
\bar x')$.  This implies that all terms cancel out except the 
phasepoint density
of the shooting point
\begin{eqnarray}
P_{\rm acc}[{\bf x}^{\rm (o)} \rightarrow {\bf x}^{\rm (n)}]  = 
\hat{h}({\bf x}^{\rm (n)}) 
\min\left[ 1, \frac{\rho(x^{(n)}_{\rm shoot})}{
\rho(x^{(o)}_{\rm shoot})
} \right]
\label{accTPS}
\end{eqnarray}
This is very convenient as this acceptance/rejection step can take place 
before the expensive trajectory generation takes place. Still, 
some (partly) completed trajectories will be rejected in the end due to the 
condition $\hat{h}({\bf x})$. However, as the new trajectory was generated 
from a small modification of an existing trajectory with $\hat{h}({\bf x})=1$ the chances 
are relatively high that the condition will be satisfied for the trial trajectory as well.   

The sampling of trajectories under a given condition $\hat{h}$ might benefit
from using a path ensemble that has a non-fixed length. 
Using a fixed path length to sample all possible trajectories between $A$ and $B$ is expensive as this length needs to be adapted  
to the longest pathway connecting these states.
 Many trajectories will reach $A$ to $B$ in a much shorter time
and will, therefore, consist of unnecessary parts that  
are not relevant for the actual barrier crossing event. 
In addition, if trajectories have significant parts outside the barrier region, the shooting move becomes inefficient as many shooting points will lie inside
the reactant or product well. Shooting from these points gives a very low probability to connect
both states. 
Using flexible path lengths was 
first introduced in Ref.~\cite{ErpMoBol2003} within the context of the 
TIS rate evaluation. However, also for the generation of reactive trajectories,
the flexible path ensemble is very useful and allows to generate paths
that start and end just at the boundaries of $A$ and $B$ 
(see Fig.~\ref{figTPS}). The only difference with the previous example is that 
$P_{\rm sel}=1/n$ is not cancelled as the trajectory length can be different.
Therefore, if the shooting procedure selects 
the timeslices by an equal probability,
the acceptance rule becomes 
\begin{eqnarray}
P_{\rm acc}[{\bf x}^{\rm (o)} \rightarrow {\bf x}^{\rm (n)}]  = 
\hat{h}({\bf x}^{\rm (n)}) 
\min\left[ 1, \frac{\rho(x^{(n)}_{\rm shoot})}{
\rho(x^{(o)}_{\rm shoot}) }
\frac{n^{(o)}}{n^{(n)}}
 \right]
\end{eqnarray}
with $n^{(o)}, n^{(n)}$ the length of the old and new path. This expression
is not so convenient as a rejection can only be made whenever the whole path is completed. Hence, the integration needs to be carried out even if 
$\rho(x^{(n)}_{\rm shoot}) << \rho(x^{(o)}_{\rm shoot})$ implying an almost certain rejection. We can, however,  separate the acceptance into two
steps by writing
\begin{eqnarray}
P_{\rm acc}[{\bf x}^{\rm (o)} \rightarrow {\bf x}^{\rm (n)}]  = 
\hat{h}({\bf x}^{\rm (n)}) 
\min\left[ 1, \frac{\rho(x^{(n)}_{\rm shoot})}{
\rho(x^{(o)}_{\rm shoot})} \right] \times \min\left[ 1,
\frac{n^{(o)}}{n^{(n)}}
 \right]
\end{eqnarray}
This acceptance rule obeys detailed balance as well and allows to reject
a modification of the shooting move that gives a too high energy. 
Still, even if the first step is accepted, the final trajectory
might be rejected whenever it becomes too long compared to the previous path.
We can improve the efficiency even further using following
trick~\cite{ErpMoBol2003}.
Instead of taking a random number $\alpha \in [0:1]$ after finishing 
our trajectory and then accept if $\alpha < n^{(o)}/n^{(n)}$, we will actually
draw this random number before starting the integration of motions. As we now 
know that we will  have to reject 
our trajectory whenever $\alpha < n^{(o)}/n^{(n)}$,
we can simply define a maximum allowed path length of this trial move in advance
\begin{eqnarray}
n^{\rm max}= {\rm int}[n^{(o)}/\alpha]
\end{eqnarray}
This allows to directly stop our trial move whenever it exceeds this maximum 
path length.

The original TPS method also provided an algorithm to determine the reaction rate of the 
process. This approach has been improved 
by the TIS~\cite{ErpMoBol2003} and RETIS~\cite{vanErp07PRL,RETIS2} algorithms. 
Like RF, the TPS rate evaluation does require a RC (I will not make 
the distinction between orderparameter or RC). However, one can show that,
compared to the RF method,
the efficiency of  
TPS, TIS, and RETIS,
is less sensitive  to an improper choice of the RC~\cite{TISeff}.

The original TPS rate evaluation is based on following correlation function
\begin{equation} 
C(t)= \frac{ \left \langle h_A(x_0) h_B(x_t) \right \rangle }{\left
\langle h_A(x_0) \right \rangle }.
\label{corrTPS}
\end{equation}
where 
$h_{A/B}(x) = 1$  if $x \in A/B$ and 0 otherwise. 
Just like Eqs.~\ref{BC}, $C(t)$ will initially show some oscillations. However,
if there is a
separation of timescales, this correlation
function grows linearly in time, $C(t) \sim k_{AB} t$, for times $\tau_{\rm mol} <t <\tau_{\rm rxn}$.
Hence,
\begin{equation} \tilde{k}_{AB}(t)=
\frac{\mathrm{d}}{\mathrm{d}t} C(t)=
\frac{ \left \langle h_A(x_0) \dot{h}_B(x_t) \right \rangle }{\left
\langle h_A(x_0) \right \rangle }, \quad
k_{AB}=\tilde{k}_{AB}(t') \textrm{ for }  \tau_{\rm mol} < t < \tau_{\rm rxn}
\label{rateTPS}
\end{equation}
The correlation function $C(t)$ is calculated in the TPS scheme using the 
shooting algorithm in combination with 
umbrella sampling. 
First, the fixed path length $t'$ is fixed to a value where $C(t)$ should give a plateau.
Then a series of path sampling simulations will be performed in 
which the final region $B$
is slowly shrunk in successive steps from the entire phase space  to the final stable state
$B$~\cite{Dellago02}. For each step numerous trajectories are generated with that condition
that the path should start in $A$ and end in the extended region $B$ at time $t'$. 
The distribution of the path's end-point  will be binned into histograms 
that will be matched 
just like ordinary umbrella sampling. Once the fully matched histogram is obtained, $C(t)$ is obtained by integration of this histogram over the actual region $B$.

The approach is rather time-consuming because
 it can take a relatively long
time $\tau_{\rm mol}$ before $C(t)$ reaches a plateau
(longer than in a transmission coefficient calculation \cite{Dellago02}). 
In Ref.~\cite{TPS99} an improvement of this approach was presented in which the umbrella sampling series could be performed with paths shorter than $ \tau_{\rm mol}$. The results were then corrected by a factor that is obtained from a single path sampling simulation using 
the longer paths. Unfortunately, the relative error in this correction factor is large if
the path length is reduced too much, so that the gain in CPU efficiency remains 
limited~\cite{ErpMoBol2003}.
Moreover, inspection of Eqs.~(\ref{corrTPS}) and
(\ref{rateTPS}) shows that a necessary
cancellation of positive and negative terms
can slow down the convergence of the MC sampling procedure.

\section{Transition Interface Sampling}
TIS is a more efficient way to calculate reaction constant than the method discussed above. The TIS methodology is also the basis of several other 
algorithms~\cite{MoBolErp2004,FFS,vanErp07PRL,Wenzel,Jutta} of which 
the PPTIS, RETIS, and FFS methods will be discussed in forth-coming sections.
The TIS rate equation is related to both the EPF expression, Eq.~\ref{kepf},  and to the  correlation
function used in TPS, Eq.~\ref{corrTPS}, albeit using different kind of characteristic functions.
Instead of using the characteristic functions of  the stable states $A$ and $B$,  we 
will redefine the correlation function using \emph{overall states} $\mathcal A$ and $\mathcal B$.
These states do not only depend on the position  at the time of consideration but also on its past behavior. Overall state $\mathcal A$ covers all phase space points lying inside stable region $A$, which constitutes the largest part, but also all phase space points that visit $A$, before reaching $B$ when the equations of motion are integrated backward in time. In other words, all phasepoints that were more recently in $A$ rather than in $B$. Similarly, state $\mathcal B$ comprises stable state B and all phase points, coming directly from this state in the past, i.e. with- out having been in $A$.
The corresponding correlation function is
\begin{equation} 
C(t)= \frac{ \left \langle h_{\mathcal A}(x_0) h_{\mathcal B}(x_t) \right \rangle }{\left
\langle h_{\mathcal A}(x_0) \right \rangle }.
\label{corrTIS}
\end{equation}
where $h_{\mathcal A}(x_0)=h^b_{AB}(x_0)$ and $h_{\mathcal B}(x_0)=h^b_{BA}(x_0)$.
Contrary to Eq.~\ref{corrTPS}, this correlation function has no oscillatory behavior during
a molecular timescale $\tau_{\rm mol}$. On the contrary,
it exhibits a linear regime 
$\sim k_{AB} t$ for $0 < t < \tau_{\rm rxn}$. 
The system will only transfer from overall state $\mathcal A$ to overall state $\mathcal B$ when
it enters region $B$ for the first time since it left region $A$. If it leaves state $B$ shortly thereafter, it will remain in $\mathcal B$. Therefore $h_{\mathcal B}(x_t)$ and  $h_{\mathcal A}(x_t)$
do not show the fast fluctuations that 
are found for $h_{B}(x_t)$ and  $h_{A}(x_t)$.
As Eq.~\ref{corrTIS} is linear from the start,  we can simply take the time derivative at $t = 0$,
which gives 
\begin{eqnarray} 
k_{AB}&=& \frac{ \left \langle h_{\mathcal A}(x_0) \dot{h}_{\mathcal B}(x_0) \right \rangle }{\left
\langle h_{\mathcal A}(x_0) \right \rangle }= 
\lim_{\ud t \rightarrow 0} 
\frac{1}{\ud t}
\frac{ \left \langle 
h^b_{AB}(x_0) 
\theta( \lambda_B-\lambda(x_{0}) )
\theta( \lambda(x_{\ud t})-\lambda_B )
 \right \rangle }{\left
\langle h_{\mathcal A}(x_0) \right \rangle } \nonumber \\
 &=&
 \frac{ \left \langle 
h^b_{AB}(x_0) 
\dot{\lambda}\left(x_0\right) 
\delta( \lambda(x_{0})-\lambda_B )
  \theta\left(\dot{\lambda}\left(x_0\right)  \right)
 \right \rangle }{\left
\langle h_{\mathcal A}(x_0) \right \rangle }
\label{kTIS}
\end{eqnarray} 
The resulting expression is basically the EPF expression (Eq.~\ref{kepf}) through the interface $\lambda_B$. 
At a first sight it might seem that generating long trajectories is no longer 
needed. As we only need $\ud C(t)/\ud t$ at $t=0$, the minimum 
time range over which
we need to calculate $C(t)$ is $[0:\ud t]$ instead of $[0:\tau_{\rm mol}]$. 
Unfortunately, unlike $h_{A}(x_0)$ and
$h_{B}(x_0)$, the determination of $h_{\mathcal A}(x_0)$ and
$h_{\mathcal B}(x_0)$ can not be done instantaneously. For this we still need
to integrate the equations of motion. However, for most
$x_0$, $h_{\mathcal A}(x_0)$/$h_{\mathcal B}(x_0)$ can be assigned 1 or 0 using a much shorter 
backward trajectory than $\tau_{\rm mol}$. For stochastic dynamics 
$h_{\mathcal A}(x_0)$/$h_{\mathcal B}(x_0)$
can be, strictly speaking, a fractional number. However,
there is generally no need to know this fractional number for a specific 
phasepoint, except for committor analysis~\cite{Ryter,Peters,vdE1,vdE2,revTPS2009}. Hence, TIS
algorithms will generally compute $h_{A}(x_0)$/$h_{B}(x_0)$ for one specific 
path to which $x_0$ belongs. Conceptually, it is therefore more accurate to 
speak of a MC sampling in pathspace rather than phasespace. 
The TIS correlation function has an additional advantage that the reaction 
rate is somewhat better defined if the separation
of timescales $\tau_{\rm mol} \ll \tau_{\rm rxn}$ is not sufficiently obeyed.
The fact that the derivative of $C(t)$ is taken at $t=0$ makes 
corrections like the one suggested in 
Ref.~\cite{impRF} unnecessary.

The TIS algorithm expresses the rate equation, Eq.~\ref{kTIS}, 
as a product of different terms. Each term has a much higher value
than the final rate and is, therefore, much easier to compute.
To introduce the TIS and PPTIS expression, that I will 
discuss in the next section, it is convenient to introduce following 
crossing probabilities that depend on four non-intersecting interfaces
$\{x|\lambda(x)=\lambda_i\}, \{x|\lambda(x)=\lambda_j\},\{x|\lambda(x)=\lambda_k\},\{x|\lambda(x)=\lambda_l\}$
\begin{eqnarray}
P(^k_l|_i^j) &=&
\lim_{\ud t \rightarrow 0}  
\frac{
\lo
h^b_{ij}(x_0) \theta(\lambda_j -\lambda(x_0))  \theta( \lambda(x_{\ud t})-\lambda_j) h^f_{kl}(x_0)  
\rc
}{
\lo
h^b_{ij}(x_0) \theta(\lambda_j -\lambda(x_0))  \theta( \lambda(x_{\ud t})-\lambda_j) 
\rc
} \nonumber \\
&= &
\frac{
\lo
h^b_{ij}(x_0) 
\dot{\lambda}\left(x_0\right) 
\delta\left(\lambda\left(x_0\right) -\lambda_j \right) 
\theta\left(\dot{\lambda}\left(x_0\right)  \right)
h^f_{kl}(x_0)  
\rc
}{
\lo
h^b_{ij}(x_0) 
\dot{\lambda}\left(x_0\right) 
\delta\left(\lambda\left(x_0\right) -\lambda_j \right) 
\theta\left(\dot{\lambda}\left(x_0\right)  \right)
\rc
} \textrm{ for } \lambda_j > \lambda_i
\label{Pcross}
\end{eqnarray}
For $\lambda_j < \lambda_i$ we simply need to replace 
$\theta(\lambda_j -\lambda(x_0))  \theta( \lambda(x_{\ud t})-\lambda_j)$
by
$\theta(\lambda(x_0)-\lambda_j)  \theta( \lambda_j-\lambda(x_{\ud t}))$
in the first line or 
$
\dot{\lambda}\left(x_0\right) 
\delta\left(\lambda\left(x_0\right) -\lambda_j \right) 
\theta\left(\dot{\lambda}\left(x_0\right)  \right)$
by
$
-\dot{\lambda}\left(x_0\right) 
\delta\left(\lambda\left(x_0\right) -\lambda_j \right) 
\theta\left(-\dot{\lambda}\left(x_0\right)  \right)$
in the second line of the above definition. 
Eq.~\ref{Pcross} defines a conditional crossing probability. It is the probability that the system will cross interface $\lambda_k$ before $\lambda_l$ under
a twofold condition. These conditions are that the system should cross
interface $\lambda_j$ right now at time $t=0$, while $\lambda_i$ was 
more recently crossed than $\lambda_j$ in the past.

\begin{figure}[ht!]
  \begin{center}
  \includegraphics[width=11.1cm]{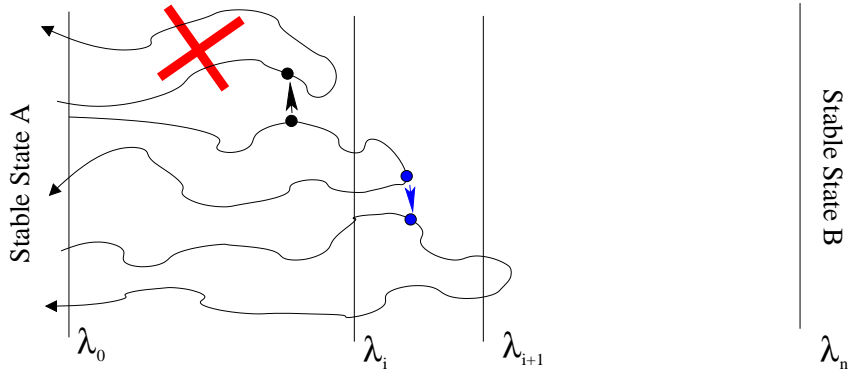}
   \caption{The TIS path ensemble $[i^+]$ is required to calculate
the conditional crossing probability ${\mathcal P}_A(\lambda_{i+1}|\lambda_i)$.
For this purpose we apply the shooting move to generate all possible 
trajectories starting at $\lambda_0$ and ending at  $\lambda_0$ or  $\lambda_n$
with at least one crossing with  $\lambda_i$. Suppose the algorithm starts with the middle  path that already fulfills these requirements. A shooting point is randomly selected and modified (black dots). However, the trajectory that
starts out from this point (top trajectory) fails to cross $\lambda_i$ and is therefore rejected and the old path is counted again. A new shooting (blue dots)
generates a valid trajectory that not only crosses  $\lambda_i$ but
$\lambda_{i+1}$ as well. This trajectory is called "successful". The fraction of
successful trajectories in this ensemble determine ${\mathcal P}_A(\lambda_{i+1}|\lambda_i)$.}
\label{figTIS}
  \end{center}
\end{figure}

Using these crossing probabilities, one can prove that Eq.~\ref{kTIS} 
is equivalent 
to the product of the initial flux times the overall 
crossing probability~\cite{ErpMoBol2003}
\begin{eqnarray}
k_{AB}=
\frac{
\lo
\dot{\lambda}\left(x_0\right) 
\delta\left(\lambda\left(x_0\right) -\lambda_0 \right) 
\theta\left(\dot{\lambda}\left(x_0\right)  \right)
\rc
}{
\lo
h_{\mathcal A}(x_0) 
\rc
} \times P(^n_0|^0_{0^{-}}) \equiv f_A {\mathcal P}_A(\lambda_n|\lambda_0)
\label{fAP}
\end{eqnarray}
where $\lambda_0$ and $\lambda_n$ are the boundaries of the stable 
states $A$ and $B$.
$f_A$ is just the flux out of state $A$ that can be computed with standard MD
as the boundary of $A$ is set at the left side of the barrier region. 
The minus in $0^-$ is to denote an interface $\lambda_0-\epsilon$ 
which is put there to indicate the direction of the crossing at $t=0$.
The overall crossing probability 
${\mathcal P}_A(\lambda_n|\lambda_0)=P(^n_0|^0_{0^{-}})$ is the probability that once $\lambda_0$ is crossed, $\lambda_n$ will be crossed before a 
recrossing with $\lambda_0$ occurs. This probability is very small, 
but it can be calculated by defining $n-1$ non-intersecting interfaces 
in between 
$\lambda_0$ and $\lambda_n$ and express the overall crossing probability as 
the following product~\cite{ErpMoBol2003}
\begin{eqnarray}
{\mathcal P}_A(\lambda_{n}|\lambda_0)= 
\prod_{i=0}^{n-1} {\mathcal P}_A(\lambda_{i+1}|\lambda_i)
\label{Pocross}
\end{eqnarray}
The factorization of ${\mathcal P}_A(\lambda_{n}|\lambda_0)$
into
probabilities ${\mathcal P}_A(\lambda_{i+1}|\lambda_i)$ that are much
higher than the overall crossing probability, is the basis of the importance
sampling approach.  
It is important to note that ${\mathcal P}_A(\lambda_{i+1}|\lambda_i)$ are
in fact complicated history dependent conditional probabilities. 
If we
consider all possible pathways that start 
at $\lambda_A$ and end by either crossing $\lambda_A$ or $\lambda_B$, while 
have at least one crossing with $\lambda_i$ in between, the fraction that
crosses $\lambda_{i+1}$ as well equals 
${\mathcal P}_A(\lambda_{i+1}|\lambda_i)$. 
This basically reduces the problem to a correct sampling of trajectories
that should obey the $\lambda_i$ crossing condition (See fig.~\ref{figTIS}).
From now on we will call this the $[i^+]$ path ensemble.
In TIS, this is done via the shooting algorithm for flexible path lengths as
is discussed in previous section (For a full flowchart diagram of the TIS algorithm see Ref.~\cite{vanErpPCCP}). 
The number of interfaces and their separation should be set to maximize 
efficiency. In Ref.~\cite{ErpBol2005,TISeff} it was found that the optimal interface
separation is obtained when one out of five trajectories reach the next
interface. 
In addition, one can define a set of sub-interfaces of arbitrary separation 
in order to construct the crossing probability as a continuous function. 
This strictly decreasing function could be viewed as the dynamical 
analogue of the free energy profile $F(\lambda)$.

There are some small differences how to treat  the path ensemble $[i^+]$ 
regarding the end-point of the path. In the first TIS algorithms, the trajectory
could reach upto $\lambda_{i+1}$ where the trajectory was stopped and assigned successful. More recent simulations continue the trajectory until reaching the stable states $A$ or $B$ each time. The additional cost is very limited as about 80\% is not reaching
$\lambda_{i+1}$ and need to be followed until reaching $A$ anyway. The choice 
to continue the trajectory even after $\lambda_{i+1}$ has the advantage that 
one can start the $[i^+]$ path ensemble without the need to fix a value for 
$\lambda_{i+1}$  beforehand. After some simulation cycles the $\lambda_{i+1}$
can be set to have the optimal 20\%  success-rate after 
which one can start
the $[(i+1)^+]$ path ensemble. In addition, the new approach makes it much more easy to use replica exchange which we will discuss in Sec.~\ref{secRETIS}. 

The simplicity of Eq.~\ref{Pocross} is deceptive and could be mistaken 
as a 
Markovian approximation. The reason that the equation is still exact lies in 
the fact that the crossing probabilities are history dependent and by the fact
that it only considers first crossing events.
We can argue the exactness of the equations also in another way.
Suppose we want to calculate the probability to go from
$\lambda_0$ to $\lambda_1$  to  $\lambda_2$ \ldots to $\lambda_n$ in successive
jumps. This probability can be expressed  as
\begin{eqnarray}
P(\lambda_0\rightarrow \lambda_1 \rightarrow \lambda_2 \rightarrow \ldots
\rightarrow \lambda_n )
&=&
P(\lambda_0) 
\times 
P(\lambda_0 \rightarrow \lambda_1|\lambda_0) 
\times
P(\lambda_1 \rightarrow \lambda_2| \lambda_0 \rightarrow \lambda_1 ) 
\nonumber \\
&\times&
P(\lambda_2 \rightarrow \lambda_3| 
\lambda_0 \rightarrow \lambda_1 \rightarrow \lambda_2)  \times \ldots
\nonumber \\
&\times&  
P(\lambda_{n-1} \rightarrow \lambda_{n}
| \lambda_0 \rightarrow \lambda_1 \rightarrow  \lambda_2  \rightarrow \ldots
\lambda_{n-2} \rightarrow \lambda_{n-1})
\end{eqnarray}
This is an exact non-Markovian expression for this specific crossing sequence 
that looks similar to Eq.~\ref{Pocross}.
However, it doesnot say anything about the many different trajectories
that could connect $\lambda_0$ with  $\lambda_n$. For instance, we should also
take into account the sequence $\lambda_0 \rightarrow \lambda_1 \rightarrow  \lambda_2 \rightarrow  \lambda_1 \rightarrow   \lambda_2  \rightarrow  \lambda_3 \rightarrow \ldots \lambda_{n-1} \rightarrow \lambda_{n}$.
Therefore, it might seem that the right expression should look much more complicated than Eq.~\ref{Pocross}. The trick, however, is that this last sequence 
can not occur if we only consider first crossing events. When we move back to $\lambda_1$ in the third step, this move will simply not considered as it is a second visit since leaving $\lambda_0$. Hence,  the successive
sequence 
$\lambda_0 \rightarrow \lambda_2  \rightarrow  \lambda_3 \rightarrow \ldots 
\lambda_{n-1} \rightarrow  \lambda_n$ is the only 
possible sequence of first crossing events that brings you from $\lambda_0$ to
$\lambda_n$.

\section{Partial Path Sampling}
The PPTIS is a variation of the TIS algorithm that was devised to 
treat diffusive barrier crossings~\cite{MoBolErp2004}. 
Despite the existence of a fine separation of timescale, 
i.e. 
the time to cross the barrier is still negligible compared to the 
time spend in the reactant well,
the path length can become too long for an effective 
computation of the reaction rate. This is the case if the barriers are 
sufficiently high to ensure exponential relaxation, but not very sharp so 
that system can move backward and forward on the barrier before it eventually drops off. 
The PPTIS equation depends on the same rate equation as TIS
\begin{eqnarray}
k_{AB}=f_A P(^n_0|^1_0)
\end{eqnarray}
The only difference with Eq.~\ref{fAP} is that we now consider the condition
$|^1_0)$ instead of $|^0_{0^{-}})$, but this is just a technical detail. 
If we take $\lambda_1=\lambda_0+\epsilon$ the equations become equivalent.
However, $\lambda_1$ can be any value that is somewhat larger than $\lambda_0$
by redefining $f_A$  as the effective flux through $\lambda_1$. This implies 
that we should count the positive crossings with $\lambda_1$ whenever the
system leaves the stable state $A:\{x|\lambda(x)<\lambda_0\}$. However, the 
next positive crossing should only be counted if the system has revisited
$A$ again. The PPTIS approach tries to avoid the generation of very long trajectories using a soft Markovian approximation. The PPTIS scheme assumes that 
for a well positioned set of interfaces the system will lose its memory
over a distance that is similar to the interface separation.
This implies for any $m>1$
\begin{eqnarray}
P(^l_k|^j_{j\pm m}) \approx P(^l_k|^j_{j\pm 1}) 
\label{Markov}
\end{eqnarray} 
The PPTIS algorithm 
consist again of a series of path sampling simulations.
Each PPTIS simulation samples a certain path ensemble in which
trajectories are confined within
two next-nearest interfaces. For instance, the $[i^\pm]$ path ensemble will
consist of all possible trajectories starting and ending at either
$\lambda_{i+1}$ or $\lambda_{i-1}$ having at least one crossing with the middle interface $\lambda_i$. From these simulations the following short-distance crossing can be obtained
\begin{eqnarray}
p_i^\pm\equiv P(^{i+1}_{i-1}|_{i-1}^i),  \quad
p_i^=\equiv P(^{i-1}_{i+1}|_{i-1}^i),  \quad
p_i^\mp\equiv P(^{i-1}_{i+1}|_{i+1}^i), \quad
p_i^\ddagger \equiv P(^{i+1}_{i-1}|_{i+1}^i)
\end{eqnarray}
with $p_i^\pm+p_i^= =p_i^\mp+p_i^\ddagger=1$.
For instance, $p_i^\pm$ is determined by dividing 
the number of trajectories in the $[i^\pm]$ ensemble that start at 
$\lambda_{i-1}$ and end at $\lambda_{i+1}$ divided by all trajectories that
start at $\lambda_{i-1}$.

Once these short distance crossing probabilities are obtained with sufficient 
accuracy, the overall crossing probability can be obtained. 
One way to do this is to use these probabilities as input for
a kinetic MC simulation~\cite{G78}. 
However,  this is not needed for a one-dimensional RC which allows 
an elegant analytical
treatment~\cite{MoBolErp2004}. Naturally, $P(^2_0|^1_0)=p_1^\pm$, but 
the calculation of
$P(^3_0|^1_0)$ requires
already to sum up the trajectories
$0\rightarrow 1 \rightarrow 2 \rightarrow 3$, 
$0\rightarrow 1 \rightarrow 2 \rightarrow 1 \rightarrow 2 \rightarrow 3$,
etc. However, as shown in Ref.~\cite{MoBolErp2004}, one can derive 
following recursive relations to make this infinite summation
of all trajectories (include the ones of infinite length!).
These PPTIS recursive relations are the following
\begin{eqnarray}
P(^{m+1}_0|^1_0)&=&\frac{p_{m}^\pm P(^{m}_{0}|^{1}_{0})}
{p_{m}^\pm+p_{m}^= P(^0_{m}|^{m-1}_{m})}, \quad  
P(^0_{m+1}|^{m}_{m+1})
=\frac{p_{m}^\mp P(^0_{m}|^{m-1}_{m})}
{p_{m}^\pm+p_{m}^= P(^0_{m}|^{m-1}_{m})} 
\label{PPTIS0}
\end{eqnarray}
or, by defining the long-distance crossing probabilities 
$P^+_m\equiv P(^{m}_0|^1_0), P_m^-=P(^0_{m}|^{m-1}_{m})$
\begin{eqnarray}
k_{AB}&=&f_A P^+_n \nonumber \\
P_{m+1}^+&=&\frac{p_{m}^\pm P^+_{m}}
{p_{m}^\pm+p_{m}^= P_{m}^-}, \quad  
P^-_{m+1}
=\frac{p_{m}^\mp P^-_{m}}
{p_{m}^\pm+p_{m}^= P^-_{m}}, \quad \textrm{for } m>1, \quad  P^+_1=P^-_1=1
\label{PPTIS}
\end{eqnarray}
Hence, starting from the initial conditions for  $(P^+_1,P^-_1)=(1,1)$, one can successively
solve $(P^+_2,P^-_2), (P^+_3,P^-_3),\ldots, (P^+_n,P^-_n)$  
via Eq.~\ref{PPTIS}. It is important to note that
$P^-_j$ is not exactly the same as $P^+_j$ in the reverse direction. Only for $j=n$ these two probabilities can be viewed as mirror images.

Here, I will derive an alternative recursive relation that does not require the auxiliary 
reverse probabilities $P^-_j$. The derivation is similar to the one presented
in the supplemental information of Ref.~\cite{Aerts} which treats the 
simpler hopping process.
To achieve this, we will  bring $P(^0_{m}|^{m-1}_{m})$ in front of Eq.~\ref{PPTIS0}.
\begin{eqnarray}
 P(^0_{m}|^{m-1}_{m})&= \frac{p_{m}^\pm 
\left[P(^{m}_0|^1_0)-P(^{m+1}_0|^1_0) \right]}
{p_{m}^= P(^m_0|^1_0)}
\end{eqnarray}
or by incrementing $m$
\begin{eqnarray}
 P(^0_{m+1}|^{m}_{m+1})&= \frac{p_{m+1}^\pm 
\left[P(^{m+1}_0|^1_0)-P(^{m+2}_0|^1_0) \right]}
{p_{m+1}^= P(^{m+2}_0|^1_0)} \label{xxqq}
\end{eqnarray}
Moreover, we can write for $P(^{m+1}_0|^1_0)$
\begin{eqnarray}
P(^{m+1}_0|^1_0)&=&P(^{m}_0|^1_0) P(^{m+1}_0|^m_{m-1}) \nonumber \\
&=&P(^{m}_0|^1_0) \left( 1-P(^{0}_{m+1}|^m_{m-1}) \right) \nonumber \\
&=&P(^{m}_0|^1_0) \left( 1-P(^{0}_{m+1}|^m_{m+1})p_m^=/p_m^\mp \right) 
\label{xxq}
\end{eqnarray}
Then, we substitute Eq.~\ref{xxqq} in Eq.~\ref{xxq} and bring
$P(^{m+2}_0|^1_0)$ in front which yields
\begin{eqnarray}
P(^{m+2}_0|^1_0)&=
\frac{p_{m}^= p_{m+1}^\pm 
P(^{m}_0|^1_0) P(^{m+1}_0|^1_0)
}{\left(p_{m}^\mp p_{m+1}^= + p_{m}^= p_{m+1}^\pm  \right)
P(^{m}_0|^1_0)-p_{m}^\mp p_{m+1}^= P(^{m+1}_0|^1_0)}
\end{eqnarray}
or
\begin{eqnarray}
k_{AB}&=&f_A P^+_n \nonumber \\
P_{m+2}^+&=&
\frac{p_{m}^= p_{m+1}^\pm 
P_{m}^+ P_{m+1}^+
}{\left(p_{m}^\mp p_{m+1}^= + p_{m}^= p_{m+1}^\pm  \right)
P_{m}^+
-p_{m}^\mp p_{m+1}^= 
P_{m+1}^+
}, \quad P_{1}^+=1, \quad P_{2}^+ = p_1^{\pm}
\end{eqnarray}
In the case of a full Markovian assumption, $p_i^\pm=p_i^\ddagger$ and
$p_i^\mp=p_i^=$, we reobtain the simpler expression of Ref.~\cite{Aerts}. 
For some this new expression might seem more esthetic as the set of two 
equations
has now been transferred into a single recursive equation without relying on the auxiliary
probability $P^-_m$. The new function has its utility~\cite{Aerts}, 
but is numerically somewhat
problematic as it can sometimes produce zeros in both nominator and denominator that 
cancel, but is  not practical for numerical calculations.

\begin{figure}[ht!]
\begin{center}
\includegraphics[width=11.1cm]{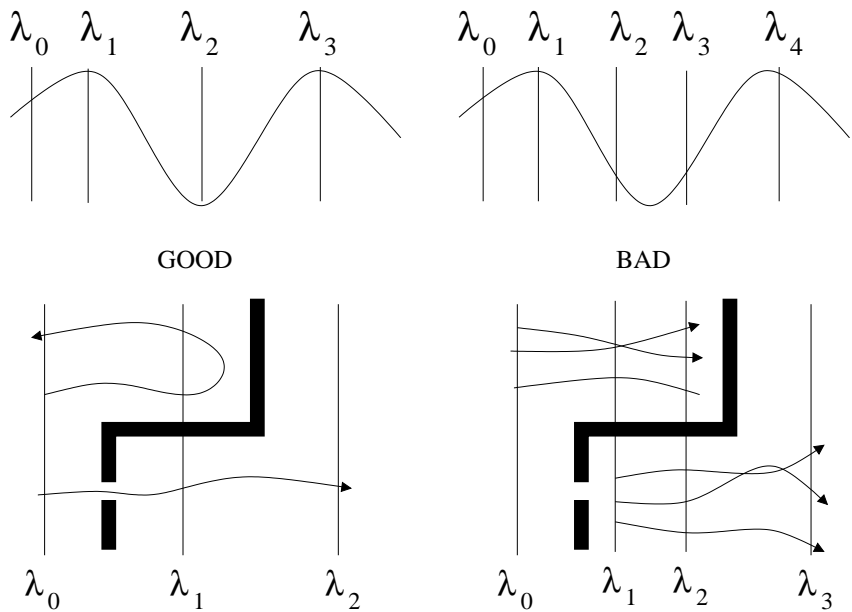}
\caption{Examples of well and badly positioned interfaces
with respect to the memory-loss assumption Eq.~\ref{Markov}.
The top situation requires a good description of the kinetic correlations
whenever the top of the barrier has many small local wells.
At the left, the interfaces are correctly placed. Once the system crosses
$\lambda_1$ it will gain a lot of kinetic energy when arriving at $\lambda_2$.
Henceforth, the chances are high that the system won't get trapped and directly moves upward to cross $\lambda_3$. The PPTIS simulation for this interface configuration will show that $p_2^\pm \gg \frac{1}{2}$ as it should. At the right-hand side we have put an additional interface inside the local well. The $[3^\pm]$
path ensemble will consists of trajectories having a much lower  
kinetic energy than the  $[2^\pm]$ ensemble of the left-hand side.
Henceforth, the right-hand side will overestimate the probability to get trapped. The bottom picture shows impermeable wall (thick black line) with a 
small hole. The left hand side shows a correct positioning of interfaces. 
The pathways that are generated from $\lambda_0$ to $\lambda_2$ all have to move through
the small hole. Conversely, the right-hand side shows a bad overlap between 
the $[1^\pm]$ and $[2^\pm]$ path ensembles which might
give the impression that trajectories can tunnel through the wall.
}
\label{figPPTIS}
  \end{center}
\end{figure}

The positioning of interfaces is crucial in PPTIS. On one hand, one would like to put them close together to improve efficiency. On the other hand,
putting them to close will introduce systematic errors due to a decrease of 
history dependence of the hopping probabilities, which invalidates 
Eq.~\ref{Markov}.
A way to measure whether the interfaces are sufficiently far is the calculation of a memory-loss function~\cite{MoBolErp2004}. However, the memory-loss
function can only 
provide a necessary but not 
necessarily sufficient condition for this separation. 
In fig.~\ref{figPPTIS} we give 
two examples of well and badly placed interfaces.

The milestoning method~\cite{FaEl2004} is very similar to PPTIS. There are basically two important differences. Milestoning assumes full memory loss
once the system hits an interface. In our notation this could be rephrased as
$P(^l_k|^j_{j\pm m}) \approx P(^l_k|^j_{j})$. This is a stronger approximation
than  Eq.~\ref{Markov}. 
The approximation of milestoning becomes exact, if the interfaces
coincide with the iso-committor functions~\cite{isocom}, but these are 
difficult determine.
On the other hand, milestoning is more
precise in the construction of the time-evolution of the system
by making the crossing probabilities time-dependent.
This is important if  there is not a clear separation of timescales 
and also allows to  calculate other dynamical properties like diffusion.
Hence, instead of $p_i^\pm, p_i^=, p_i^\mp, p_i^\ddagger$ 
with $p_i^\pm+p_i^= = p_i^\mp + p_i^\ddagger=1$, milestoning calculates for 
each interface the time-dependent probability densities $p_i^+(t), p_i^-(t)$ 
with $\int_0^\infty [p_i^+(t)+p_i^-(t)] \ud t=1$.
The strengths of PPTIS and milestoning  do not exclude each other and could be 
unified into a single method as was suggested in  Ref.~\cite{MoErpBol}.
A realization of such a method was recently published~\cite{milpptis}.

\section{Forward Flux Sampling}
\begin{figure}[ht!]
  \begin{center}
  \includegraphics[width=\textwidth]{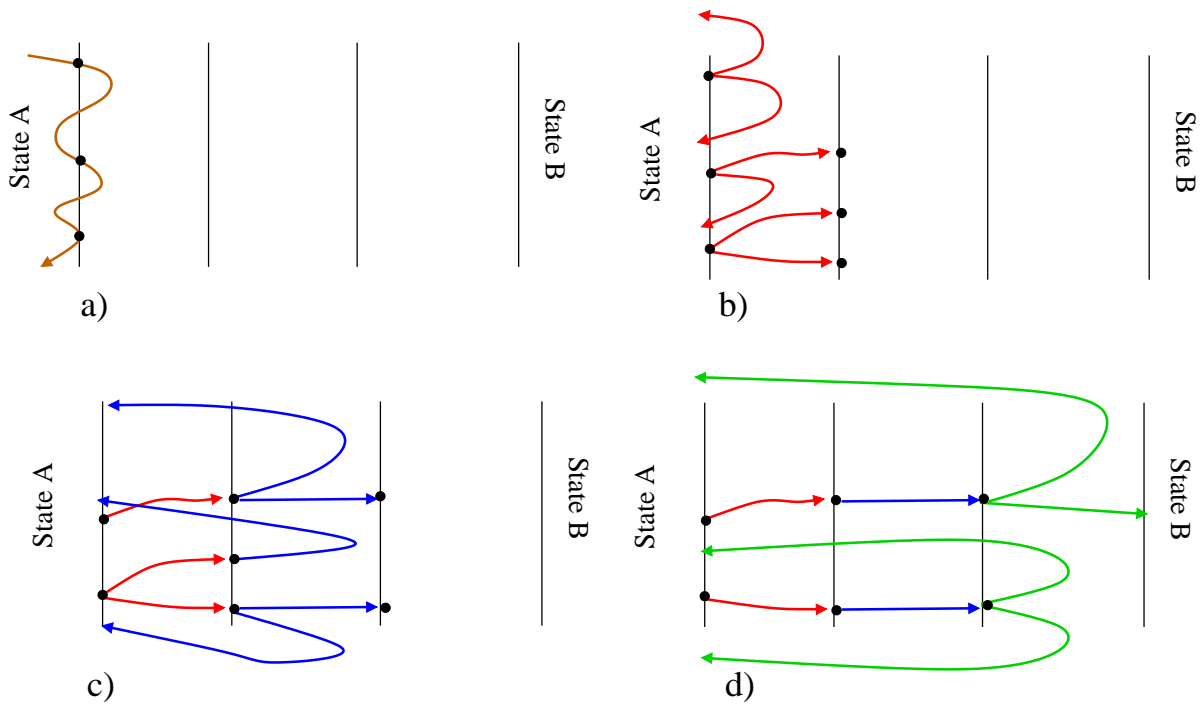}
   \caption{Illustration of FFS  algorithm. a) shows the initial MD simulation that is needed to calculate the flux $f_A$. The positive crossing points (black dots) are stored. b) Starting from the 
   stored MD crossing points a number of trajectories are released. The endpoints of the
   successful trajectories (that reach $\lambda_1$) are stored again and used for
   the next path ensemble simulation c). Finally, the reactant state will be reached d).
   \label{figFFS}}
  \end{center}
\end{figure}

FFS was originally developed for the special case of biochemical networks that do not obey
equilibrium statistics nor time-reversibility~\cite{FFS}. However, its advantageous implementation and apparent
efficiency has gained this method a fast increasing popularity for equilibrium systems as well.
FFS is based on the same  theoretical TIS rate equation~\ref{kTIS}. However, the fundamental difference is the sampling move. While the principal sampling move in TIS is the shooting move,
FFS is based on a non-Metropolis MC scheme called splitting~\cite{split1,split2}. This approach 
requires stochastic dynamics, although it has been suggested that FFS is able to treat
deterministic dynamics utilizing the Lyapunov 
instability~\cite{FFSMD1,FFSMD2} using small 
'invisible' stochastic noises~\cite{bolhuis03}. Like TIS, 
FFS consists of a straightforward MD simulation, from which
the escape flux $f_A$ is obtained, followed by a series of path sampling simulations.
However, besides giving the flux value, the MD simulation also provides the starting conditions for the path sampling simulations. Each time that the first interface $\lambda_0$ is crossed in
the positive direction, this phasepoint just after the interface is stored on the hard-disk.
 In the first path simulation, performing the $[0^+]$ path ensemble,
 these points are used as starting points for the trajectories that are continued until reaching 
 $\lambda_1$ or returning to $\lambda_0$. Naturally, stochasticity is 
of  eminent importance, otherwise all these trajectories would just reproduce parts of MD simulation. The endpoints of the trajectories
 that successfully reach  $\lambda_1$  are stored again and serve as initial points for the $[1^+]$ 
 ensemble. The path ensembles are executed one after the other  by the same procedure until 
 reaching state $B$ (See fig.~\ref{figFFS}). 
 
 There are advantages and disadvantages compared to the TIS algorithm.
 The most important advantage of FFS is that it allows to treat non-equilibrium systems as it does not require any knowledge about the
 phasepoint density. TIS 
 employs the shooting move that requires to know $\rho(x)$ for the acceptance, Eq.~\ref{accTPS}. 
 In addition, FFS does not require any integration of motion backward in time. Therefore, time-reversibility is not required. Moreover, unlike PPTIS, FFS is,
in principle,
equally exact as TIS.
 However, if one has to chose between TIS and FFS for equilibrium dynamics, one has to consider following points. FFS will generally create more trajectories for the same number
 of MD steps as it recycles
 previously generated trajectories. Moreover, there are 
no rejections like there are in TIS and any other Metropolis based MC scheme. 
In practice, the reduction in MD steps will be limited to a certain 
factor ($\approx 2,3$)
as the unsuccessful trajectories, which is the largest part,
have to be followed until reaching  state $A$.
 On the other hand, the FFS trajectories will be  much more 
 correlated than the TIS trajectories. 
 This implies that FFS needs much more trajectories than TIS to obtain 
the same accuracy.
 One reason for this, 
is that FFS generates several trajectories having the same starting point. Absence of stochasticity will result that all these 
 trajectories basically coincide. Fully Brownian motion does not exclude 
 correlation effects either as the successful trajectories starting 
from the same point will hit the next interface in a confined region. The
size of this region  is determined by the diffusion orthogonal to the RC and the time
 it takes to go from interface to the other. Besides correlations within 
a certain path ensemble,
 the FFS method also introduces correlations between the different 
ensembles. This is a crucial difference with TIS where the MD simulation and all path simulations are
 independent. 
One of its consequences is that
the FFS is more sensitive to the RC than TIS or even RF~\cite{TISeff}. 
An efficiency analysis of FFS~\cite{FFSeff} ignores this correlation 
effect.  This can be a rather crude approximation that is probably 
only valid 
when interfaces are  approximately equal to the isocommittor surfaces. 
Suppose  $\lambda^\perp$ denotes a coordinate orthogonal to the RC. Let
 ${\mathcal P}_A(\lambda_{n}|\lambda_0;\lambda^\perp)$ be the  
 overall crossing probability from $\lambda_0$ to $\lambda_n$ starting from 
 a point  $\lambda^\perp$  on the first interface.
  Then, the full
 overall crossing probability is given by
 \begin{eqnarray}
 {\mathcal P}_A(\lambda_{n}|\lambda_0) = 
 \int \ud \lambda^\perp \, {\mathcal P}_A(\lambda_{n}|\lambda_0;\lambda^\perp) 
 \rho(\lambda^\perp|\lambda_0)
\label{Pperp}
 \end{eqnarray}
 where $\rho(\lambda^\perp|\lambda_0)$ 
is the probability density of $\lambda^\perp$ 
 on interface $\lambda_0$. 
FFS  will suffer considerably  
when the distributions $\rho(\lambda^\perp|\lambda_0)$ and
${\mathcal P}_A(\lambda_{n}|\lambda_0;\lambda^\perp)$ are not overlapping. 
 In that case,  FFS will miss important crossing points that are significant 
for the rate evaluation. Some studies have shown that FFS can significantly
underestimate reaction rates~\cite{Jarek,Shear}  in practical cases.
Sampling artefacts like this, are also not yet  fully excluded 
as possible explanation 
for 
some surprising  results on
non-equilibrium nucleation~\cite{Sanz}. 

 This issue  
  will be most sensitive 
 to the MD and the first interface ensembles on which 
all the further results will depend.
 If $\lambda_1, \lambda_2, \ldots, \lambda_{n-1}$ are isocommittor surfaces then
 ${\mathcal P}_A(\lambda_{n}|\lambda_i;\lambda^\perp)$ is a constant
 as function of $\lambda^\perp$ which eliminates the problem.
This is the reason that Borrero {\emph et al.} devised a FFS 
scheme in which the interfaces are repositioned on-the-fly in order
to obtain a proper RC~\cite{FFSRC}.
 
 TIS has the advantage that it can relax the history of the path via 
the backward 
 integration. Therefore, the distribution density 
 $\rho(\lambda^\perp|\lambda_0)$ can
change when considering the different path ensembles. 
TIS can give correct results even if the sampled distribution 
of starting points 
 $\lambda^\perp$ 
 of the final  $[n-1^+]$ trajectories 
 do not overlap with the initial MD crossing points~\cite{CCP}.

\section{Replica-Exchange TIS}
\label{secRETIS}
In Ref.~\cite{vanErp07PRL}, I showed that a special type of replica 
exchange~\cite{RE2,marinari92} can significantly improve the TIS efficiency
(See also Ref~\cite{RETIS2} for some extensions of this approach). A crucial difference with 
standard RE, which has also been applied to TPS~\cite{VS01}, is that the RETIS method
does not require additional simulations at elevated temperatures. Instead, 
swaps are attempted between the different
TIS path ensembles.  
For this purpose, RETIS has replaced the
initial MD simulation  by another path ensemble, called $[0^-]$, that consists of all path that start at $\lambda_0=\lambda_A$, then go in the opposite direction away from the barrier inside state $A$, and finally end at
$\lambda_0$ again. The flux is then obtained from the average path length of
the $[0^-]$ and  $[0^+]$ ensembles as follows~\cite{vanErp07PRL}. 
\begin{eqnarray}
f_A= \Big( \lo t_{\rm path}^{[0^-]} \rc 
+ \lo t_{\rm path}^{[0^+]} \rc \Big)^{-1}
\end{eqnarray}
where $\lo t_{\rm path}^{[0^-]} \rc, \lo t_{\rm path}^{[0^+]} \rc$ 
are the average path 
lengths in the $[0^-]$ and $[0^+]$ path ensembles respectively.

\begin{figure}[ht!]
  \begin{center}
  \includegraphics[width=\textwidth]{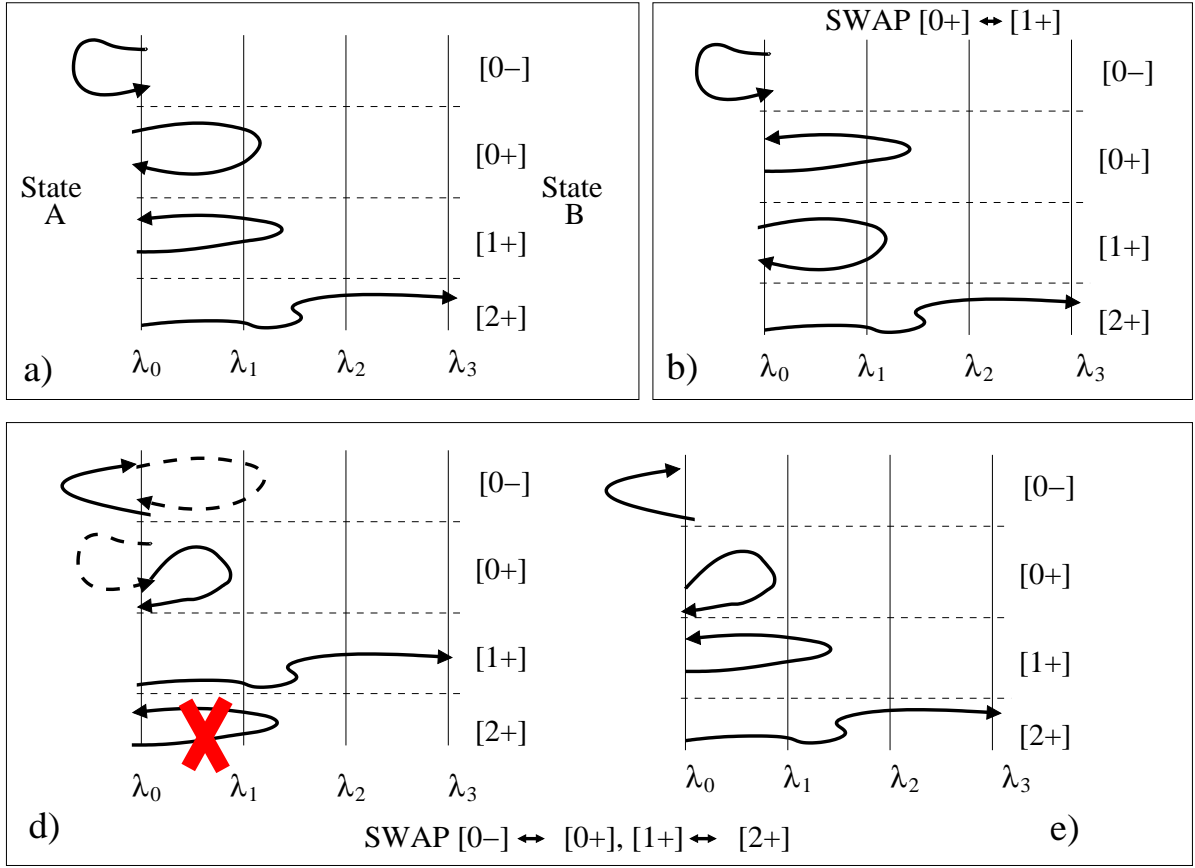}
   \caption{Parallel path swapping move in RETIS.
   The picture illustrates a RETIS simulation using four interfaces. a) shows the
   initial "superstate" that contains one trajectory per ensemble  $[0^-], [0^+], [1^+],$ and $[2^+]$.
   b) shows the superstate after the $[0^+] \leftrightarrow [1^+]$ swap. The original 
   $[0^+]$ trajectory crossed $\lambda_1$ and is therefore a valid path in the $[1^+]$ ensemble.
   The swapping move is, therefore, accepted. c) shows the trial superstate that is obtained
   after the simultaneous swaps $[0^-]  \leftrightarrow [0^+]$ and $[1^+]  \leftrightarrow [2^+]$.
   The first swap requires the integration of motion forward or backward in time starting from 
   the last of first timeslice of the swapped trajectories. This swap will 
always generate 
   acceptable trajectories for the $[0^-], [0^+]$  ensembles. The other swap $[1^+]  \leftrightarrow [2^+]$ is rejected
   because the old $[1^+]$ path does not cross $\lambda_2$. e) gives the final situation
   after the whole move.
\label{figRETIS}}
  \end{center}
\end{figure}

As the dynamical process
is now fully described by path simulations with
different interface-crossing conditions, the exchange of trajectories between them becomes extremely efficient,
especially if the process possesses multiple reaction channels~\cite{CCP}.
The methodology avoids the need of doing additional simulations at elevated temperatures and even gives  paths for free as for most swapping moves whole trajectories are being swapped.
Only when a swapping between the $[0^-]$ and $[0^+]$ ensembles are attempted,
two phase points are interchanged. From the last point
of the  $[0^-]$ trajectory a new path in the $[0^+]$ is generated. Reversely,
the first point of the old $[0^+]$ path will serve to generate
a new path in the
$[0^-]$ ensemble by integrating the equations of motion backward in time
(See Fig.~\ref{figRETIS}).

The RETIS algorithm is then as follows. At each step it is decided 
by an equal probability whether a series of shooting or swapping moves
will be performed. In the first case, all simulations will be updated 
sequentially by one shooting move. In the second case, again an equal 
probability will decide whether the swaps  $[0^-] \leftrightarrow [0^+],
[1^+] \leftrightarrow [2^+], \ldots$ or the swaps $[1^+] \leftrightarrow [2^+],
[3^+] \leftrightarrow [4^+], \ldots$ are performed. Each time that 
$[0^-]$ and $[(n-1)^+]$ do not participate in the swapping move they
are left unchanged. Also when the swapping move does not yield
valid paths for both ensembles, the move is rejected for these two simulations 
and the old paths are counted again. Note that the swapping moves 
do not require any force calculations except for the 
$[0^-] \leftrightarrow [0^+]$ swap.

Like FFS, the path ensembles in RETIS are not fully uncorrelated. However, their dependence
fundamentally different.  In FFS,  the path ensemble $[i^+]$ is fully determined by its predecessors,
the MD simulation and the path simulations $[j^+]$ with $j<i$. Conversely, the 
separate
RETIS simulations
generate a large part of their trajectories independently. The exchange between the ensembles is, therefore, an additional help instead of a strict dependence as it is for FFS. 
Moreover, the benefit of the exchange 
works in both directions and
is mutual for all ensembles, i. e. the $[i^+]$ path ensemble 
can improve the sampling in 
both $[(i+1)^+]$ and  $[(i-1)^+]$ via the swapping moves  
$[i^+]\leftrightarrow [(i-1)^+]$ and  $[i^+] \leftrightarrow [(i+1)^+]$, and will improve itself
due to the same moves.

\section{Numerical Example}
\begin{figure}[ht!]
  \begin{center}
  \includegraphics[width=\textwidth]{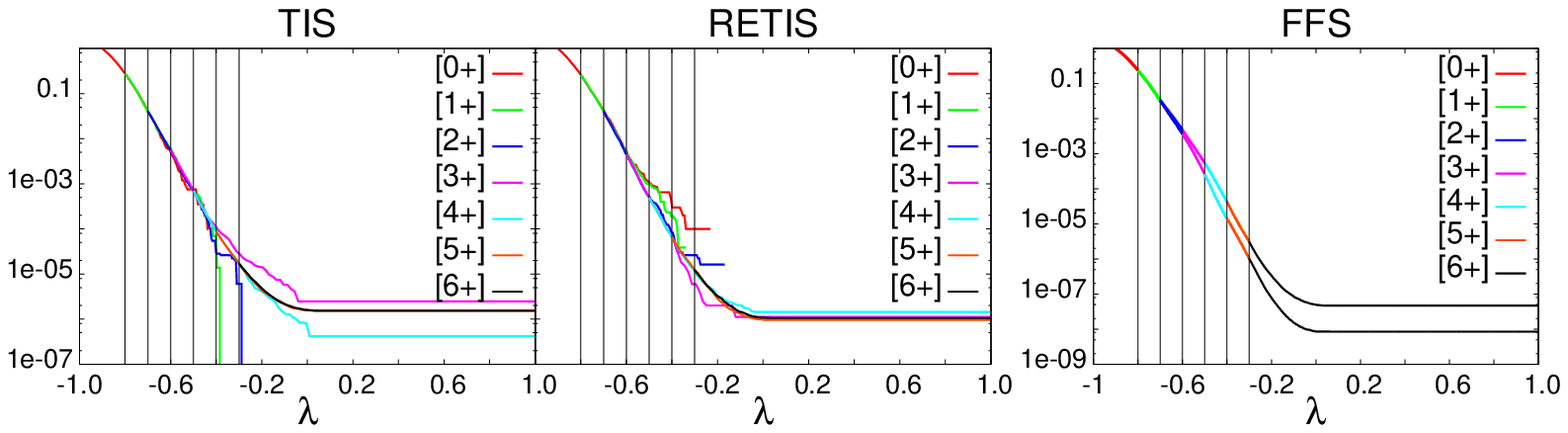}
   \caption{Overall crossing probabilities using TIS, RETIS, and FFS
\label{figTP}}
  \end{center}
\end{figure}

\begin{figure}[ht!]
  \begin{center}
  \includegraphics[width=8cm, angle =-90]{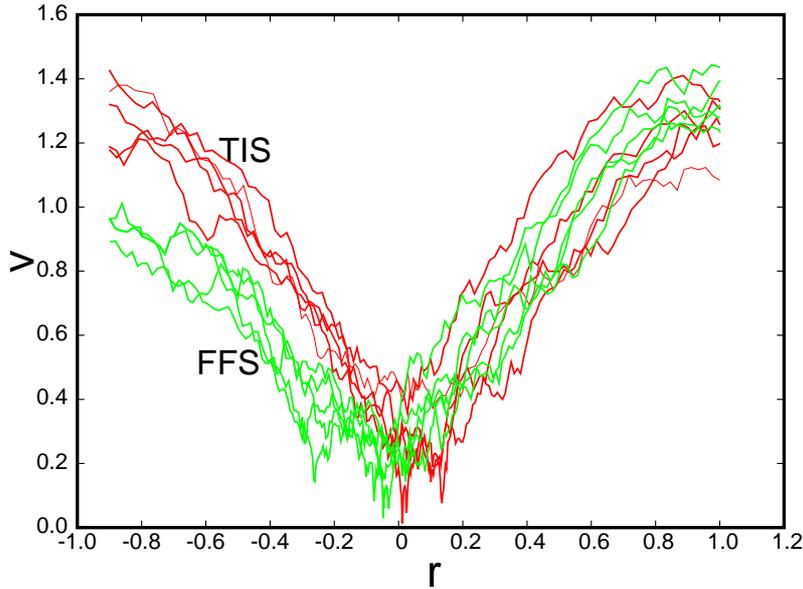}
   \caption{TIS and FFS trajectories in the $(r,v)$ plane
\label{figtraj}}
  \end{center}
\end{figure}

We will apply 
the different methods on a simple one-dimensional test system using Langevin
dynamics with finite friction. The Langevin dynamics was chosen because its 
inhibits stochasticity which is required for FFS. 
Hitherto, most studies on FFS have applied Brownian dynamics.
As the dynamics we are considering are not overdamped, the 
dimensionality of the system becomes 
effectively two-dimensional. We could consider the velocity as an  
orthogonal coordinate  
(like $\lambda^\perp$ in Eq.~\ref{Pperp}). 
This makes the choice of a proper RC not such a 
triviality as one would think at first sight. 
However, we will simply take the RC to be configurational dependent, which is
the standard approach.
The system that we will consider consist of a single one-dimensional particle
inside a double well potential
\begin{eqnarray}
V(r)=k_4 r^4-k_2 r^2
\end{eqnarray}
with $k_4=1$ and $k_2=2$. The corresponding potential 
 has a maximum at $r=0$ and two minima at $r=\pm 1$.
We use reduced units where the mass and the Boltzmann constant are set to unity, 
$k_B=m=1$. The system is coupled to a Langevin thermostat with friction coefficient $\gamma=0.3$ and temperature $T=0.07$. The equations of motion are integrated
using MD timestep of $\ud t=0.002$. 
 The RF method was applied using the EPF formalism for the transmission coefficient calculation.
 For this purpose 100,000 trajectories were released from the TST dividing surface $r=0$. 
 The free energy term 
 $\int_{-\infty}^0  \ud \lambda \, e^{-\beta F(\lambda)}$
 was obtained by a simple numerical integration. The RF method is by far the most 
 efficient method for this system because there is basically no error in the free energy calculation and
 the transmission coefficient is close to unity. The RF results will therefore be the reference for the other methods. The escape flux $f_A$ for TIS, FFS and PPTIS was determined using a MD simulation of 10,000,000 timesteps. The 
same MD result was used for these three methods. We performed an additional MD simulation using less timesteps, 4,000,000, for a second FFS calculation in order to see how this effects the final FFS result. 
I defined eight interfaces 
$\lambda_0=-0.9, \lambda_1=-0.8, \lambda_2=-0.7, \lambda_3=-0.6, 
\lambda_4=-0.5, \lambda_5=-0.4, \lambda_6=-0.3$, and $\lambda_7=1.0$.
 For each  path ensemble 20,000 trajectories were generated. For TIS and PPTIS, 50\% of the 
 MC moves were shooting moves.  I applied the  aimless 
shooting~\cite{peters06}  approach in which
the velocities at the shooting point  are completely
regenerated from Maxwellian distribution.  However, unlike Ref.~\cite{peters06}, shooting points
were picked with an equal probability for all  timeslices along the path without
considering the previous shooting point~\cite{vanErp07PRL}. 
  The other 50\% were time-reversal moves. Time-reversal moves simply change the order of the 
  timeslices of the old path while reversing the velocities. Time-reversal can sometimes
  increase the ergodic sampling and is basically cost-free as it doesnot require any force calculations. However, as aimless shooting is also able to reverse velocities in a single step, 
  the time-reversal move  could actually have been omitted for this case.  
In the RETIS algorithm there was at each step a   25\% probability to perform a shooting move, another  25\% probability to do a
time-reversal move, and a 50\% probability to do a replica exchange move. 
The FFS simulations consist of a single move which is the   forward integration of the equations of motion.  
The Langevin thermostat served for the necessarily stochasticity.
The results are shown in table \ref{tabres}. 
The RF method gives the most accurate results as expected. The value for $\kappa$
can be compared to Kramer's expression $\kappa \approx
(1/\omega_b)
(
-\gamma/2
+\sqrt{
\gamma^2/4 +\omega_b^2})
= 0.9$ with  $\omega_b=\sqrt{k_2/m}=\sqrt{2}$.
\begin{table}[htdp]
\caption{Results of the rate evaluations using RF, TIS, PPTIS, RETIS, and FFS. Final errors
were obtained by block averaging and error-propagation rules. The errors of RETIS and FFS are given a star as these errors should not be considered exact due to the neglect of 
covariant terms which arise due the correlations between path ensembles and initial  MD simulation. The FFS was repeated using a shorter (4,000,000 instead of 10,000,000 timesteps)
initial MD run. }
\begin{center}
\begin{tabular}{ccccc}
\hline 
reactive flux method & $\frac{1}{\sqrt{2 \pi \beta m}}$ &
 $\frac{e^{-\beta F(0)}}{\int_{-\infty}^0  \ud \lambda \, e^{-\beta F(\lambda)}} $ & $\kappa$ & 
 $k= 
 \kappa \times \frac{1}{\sqrt{2 \pi \beta m}} \times \frac{e^{-\beta F(0)}}{\int_{-\infty}^0  \ud \lambda \, e^{-\beta F(\lambda)}} $ 
 \\
   EPF algorithm &0.106   & $2.63\cdot 10^{-6}$ & $0.874 \pm 4 \%$  & 
$2.42 \cdot 10^{-7} \pm $ 4   \%
\\ \hline 
   &&&& \\
   \hline
    \multicolumn{2}{c} {path sampling} & $f_A$ & $ {\mathcal P}_A(\lambda_{n}|\lambda_0)$& 
    k= $f_A \times  {\mathcal P}_A(\lambda_{n}|\lambda_0)$ \\ \hline
    \multicolumn{2}{c} {TIS}    & $0.263 \pm 1\%$ &  $1.52\cdot 10^{-6} \pm 20\%$ & $4.02 \cdot 10^{-7}
\pm 20\%$ \\
    \multicolumn{2}{c} {PPTIS}    & $0.263 \pm 1\%$ &    
$1.04\cdot 10^{-6} \pm 19\%$ &
$2.73 \cdot 10^{-7} \pm 19\%$
\\
        \multicolumn{2}{c} {RETIS}    & $0.265 \pm 1\%^*$ &  
$ 1.05 \cdot 10^{-6} \pm 25\%^*   $
 & 
$2.79 \cdot 10^{-7} \pm 25\%^*   $ \\
       \multicolumn{2}{c} {FFS (long MD run)}    &  $0.263 \pm 1\%$ &  
$ 4.69 \cdot 10^{-8} \pm 6\%^* $ & 
$ 1.23 \cdot 10^{-8} \pm 6\%^* $ \\
    \multicolumn{2}{c} {FFS (short MD run)}    & $0.259 \pm 2\%$ &  
$ 8.45 \cdot 10^{-9} \pm 9\%^* $
 & $ 2.18 \cdot 10^{-9} \pm 9\%^* $ \\ \hline
\end{tabular}
\end{center}
\label{tabres}
\end{table}%
If we compare the TIS, PPTIS, and RETIS simulations we see that they are all close (within a factor 2) to the RF result. The TIS result is somewhat too high which is probably due to a single path ensemble calculation that wasn't fully converged after 20,000 steps. The RETIS results
are clearly better despite the apparent errors that are somewhat smaller for TIS. The RETIS
result is much closer to the RF reference. Moreover, it uses only halve the number of
shooting moves compared to TIS, which is the most expensive move 
for realistic systems
as it requires a large number of force evaluations.
Also, the construction of the 
overall crossing probabilities in Fig.~\ref{figTP} shows a
much better matching between the different ensembles in the RETIS method. 
The PPTIS result is also very close to the reference value. The PPTIS
approximation, Eq.~\ref{Markov}, becomes not only exact for very diffusive systems, but is also exact
for steeply increasing barriers as all trajectories from $\lambda_i$ to $\lambda_{i+1}$ come directly from
$\lambda_0$ in the past.
 FFS on the other hand, that is in principal exact unlike PPTIS,  gives an unacceptable value
 that is about a factor 20 too low. Still, if we calculate the error using standard error propagation rules without taking care of the correlations between the ensembles~\cite{FFSeff}, we 
 get errors that seem very low.  I also repeated the FFS simulation only changing the 
 length of the initial MD simulation. A decrease of 60\% for  the initial 
MD
simulation resulted in a final result that is again 5 times smaller. For TIS, PPTIS, and RETIS the impact of this MD reduction 
 would not even be noticed as it only effects the error in the  flux term that is negligible compared
 to the error in the crossing probability.  Fig.~\ref{figTP} shows that the two different FFS
 crossing probabilities  are similar at the start, but then start to deviate exponentially. The reason for this behavior is that the true set of reactive trajectories have an average kinetic energy  distribution 
at the start
that is considerably shifted compared to
 the equilibrium distribution. Therefore, a too short MD simulation might not generate
 sufficient crossing points having a high velocity. 
As result, the FFS trajectories mainly climb up the barrier helped by the stochastic force instead
of a high initial velocity. In Fig.~\ref{figtraj} we compare five randomly selected  crossing trajectories for TIS and for
FFS.  The FFS trajectories are clearly unrealistic as they are not symmetric in the $(r,v)$ plane, which should be the case for a 
symmetric barrier. The velocities at the start are much lower than at the other side of the barrier.
Moreover, two of the five trajectories, that were randomly picked from the 166 successful ones,
start exactly from the same MD crossing point (from a total of  5260 crossing points).
 TIS, that is able to relax the history of the path, does not show this 
artefact. 

These results have important consequences. It shows that attempts to use FFS 
for deterministic MD~\cite{FFSMD1,FFSMD2}
by applying some small level of stochastic noise can only work when 
inertia effects are not important. In other words, when the deterministic dynamics 
behaves effectively Brownian. 
This sampling problem is not unique to FFS, but to any splitting-type  method
such as weighted ensemble Brownian dynamics~\cite{web1,web2}, 
Russian Roulette~\cite{split1,split2}, vector walking~\cite{TNK03}, and
S-PRES~\cite{SPRES}, in which
the equations of motion are only followed forward in time. 
There is not an easy solution for this problem. 
Still, this is very much desired 
for non-equilibrium systems for which there  are no good alternatives. 
Possibly, the original nonequilibrium TPS approach of 
Crooks and Chandler~\cite{CC01} could do better in this situation as
it continues rebuilding trajectories from the beginning. Still, 
the set of starting points follows from a straightforward MD trajectory.
Therefore, also this approach is likely to miss these rare initial points
that have high potential to become reactive. Timereversal moves might alleviate 
the problem~\cite{RETIS2}, but can only be applied for time-reversible dynamics
and velocity-symmetric steady state distributions.
The other possibility is to adapt the RC, $\lambda(x)$, to the phasepoint committor,
for instance via an on-the-fly optimization scheme~\cite{FFSRC}.
Such an approach would need to be carried out  
in full phasespace, which is presently not the standard.

However, redefining the intermediate  interfaces to lie on the isocommittor surfaces alone will not be sufficient. If we keep the $r$-dependent definition 
for stable state $A$, the committor probability
 can jump from zero to one is a single timestep (at the point when it 
leaves state $A$ with a 
 high velocity). Hence, also the state definition $A$ should be redefined 
in phasespace. 
 For this particular system it seems  intuitive to use constant 
energy curves $\lambda(x)=
 1/2 m v^2+V(r)$ as RC. FFS will probably work in that case., However,
it is yet unclear if it is practically feasible  
to design  appropriate RCs using full phasespace in more complex systems. 
Present 
 algorithms~\cite{revTPS2009} have always assumed that is sufficient 
to use configuration dependent committor functions.

On the other hand, the TIS methods seems to work properly using 
configuration dependent RCs. Only to ensure the stability of state $A$
it is sometimes convenient to let $\lambda_0$ be velocity 
dependent~\cite{ErpMoBol2003} (For this system, the friction  
coefficient is sufficiently high  to neglect 
kineticly correlated recrossing). 
On the contrary, it seems that 
TIS and its variations do not necessarily  improve when the RC equals   
the true isocommittor, 
a hypothesis that was postulated   in Ref.~\cite{isocom}.
If interfaces are places at constant energy curves, the trajectories  will become much longer than in the present case.

\section{Conclusions}
I have reviewed some dynamical
rare event simulation techniques.  The RF method is likely the most efficient
approach when studying low dimensional systems for which an appropriate RC can easily be found.  The most efficient implementation of the RF approach to calculate the dynamical factor is probably the EPF algorithm that is considerably more efficient than the more common
transmission coefficient calculation schemes. However, even with this more efficient
EPF approach, the RF efficiency will decrease exponentially with barrier height and inverse temperature, if a proper RC can not be found~\cite{TISeff}.
The TPS reactive trajectory sampling does not require a RC. However, a definition of a RC is still needed in the TPS rate calculation algorithm, which has been improved by the TIS and RETIS
methodologies. The TPS/TIS/RETIS efficiency only scales quadratically
with  barrier height and inverse temperature when using an 
"improper" RC~\cite{TISeff}. The RC insensitivity
of these methods gives them a strong advantage compared to RF methods in complex systems.
Of these methods, RETIS is significantly faster than the other two. 
However, its implementation is somewhat more difficult than TIS.
PPTIS (and the similar milestoning) is not an exact method as it assumes memory loss beyond  a traveling distance between two interfaces. Using this approximation, PPTIS is able to reduce the required path length considerably, which can be important for diffusive barrier crossings.
FFS does not require information on the phasepoint density and is, therefore, ideally suited
to study nonequilibrium events. The advantageous implementation and its apparent efficiency have made FFS a very popular method for equilibrium systems as well. 
However, the  numerical study, presented here, shows that FFS has certain pitfalls that have  not yet been reported. It shows that the RC sensitivity of FFS is even more troublesome than it is for RF
methods. Present simulation studies  have almost always assumed that RC 
are functions of configuration space alone.
My example shows that an appropriate RC for FFS needs to be defined phasespace
while configurational space would be sufficient for the RF method 
and the equilibrium path sampling algorithms TPS/(RE)TIS. 
Still, there are presently no alternative methods that 
can treat
nonequilibrium processes and 
do not have the same problem. The fact 
that FFS and other forward MC methods get so  
easily trapped towards unfavorable 
reaction paths, by missing an important
orthogonal coordinate or velocity in an early stage, 
requires the uppermost caution when 
applying these methods
and interpreting their results.
In addition, this article poses challenges for developing improved nonequilibrium 
path sampling methods that are either less sensitive to a chosen RC or
are able to find an appropriate phasespace dependent RC on the fly.
\\ \\

{\bf Acknowledgments:} 
TSvE acknowledges the Flemish Government for long-term structural support
via the center of excellence (CECAT)
and Methusalem funding (CASAS) and the support of the Interuniversity Attraction Pole (IAP-PAI).

\section{List of abbreviations}

\begin{tabbing}
KWW:zzz\=\kill
EPF  \>  Effective Positive Flux \\
FFS \> Forward Flux Sampling\\
MC \> Monte Carlo \\
MD \> Molecular Dynamics \\
PPTIS \> Partial Path Transition Interface Sampling \\
RC  \> Reaction Coordinate \\
RE  \> Replica Exchange \\
RETIS  \> Replica Exchange  Transition Interface Sampling \\
RF  \> Reactive Flux method \\
TI   \> Thermodynamic Integration \\
TIS     \>Transition Interface Sampling\\
TPS 	\>Transition Path Sampling\\
TS \>Transition State \\
TST  \>Transition State Theory \\
US \> Umbrella sampling
\end{tabbing}

\bibliographystyle{unsrt}

\end{document}